\documentclass[12pt,preprint]{aastex}

\usepackage[table]{xcolor}

\newcommand{\A}{\AA \hspace{1.5mm}}
\newcommand{\grad}{${^\circ}\;$}
\shorttitle{Europa at Eastern Elongation}
\shortauthors{Saur et al.}

\begin{document}


\title{HST/ACS Observations of Europa's Atmospheric UV Emission at Eastern Elongation}

\author{Joachim Saur\altaffilmark{1},
Paul D. Feldman\altaffilmark{2}, 
Lorenz Roth\altaffilmark{1}, 
Francis Nimmo\altaffilmark{3},
Darrell F. Strobel\altaffilmark{2,4}, 
Kurt D. Retherford\altaffilmark{5}, 
Melissa A. McGrath\altaffilmark{6}, 
Nico Schilling\altaffilmark{1},
Jean-Claude G\'{e}rard \altaffilmark{7}, 
Denis Grodent\altaffilmark{7}
}

\altaffiltext{1}{Institute of Geophysics and Meteorology, University of Cologne, Germany}
\email{saur@geo.uni-koeln.de}
\altaffiltext{2}{Dept. of Physics and Astronomy, The Johns Hopkins University, Baltimore, USA}
\altaffiltext{3}{Dept. of Earth and Planetary Sciences, University of California, Santa Cruz, USA}
\altaffiltext{4}{Dept. of Earth and Planetary Sciences, The Johns Hopkins University, Baltimore, USA}
\altaffiltext{5}{Southwest Research Institute, San Antonio, USA}
\altaffiltext{6}{NASA Marshall Space Flight Center, USA}
\altaffiltext{7}{University of Li\`{e}ge, Li\`{e}ge, Belgium}

\begin{abstract}

We report results of a Hubble Space Telescope (HST) campaign 
with the Advanced Camera for 
Surveys to observe Europa at eastern elongation, i.e. Europa's leading side, 
on 2008 June 29. With five consecutive HST orbits,
we constrain Europa's atmospheric \ion{O}{1} 1304 \A  and 
\ion{O}{1} 1356  \A 
emissions  using the prism PR130L. 
The total emissions of both oxygen multiplets range between
132 $\pm$ 14 and 226 $\pm$ 14 Rayleigh. \textcolor{black}{ 
An additional systematic error
with values on the same order as the statistical errors may
be due to uncertainties in modelling the reflected light from
Europa's surface.}
The total emission also  
shows a clear dependence of Europa's position with 
respect to Jupiter's magnetospheric plasma sheet. We derive
a lower limit for the O$_2$ column density of 6 $\times$ 10$^{18}$ m$^{-2}$.
Previous observations of Europa's atmosphere with STIS 
in 1999 of Europa's trailing
side show an enigmatic surplus of radiation on the anti-Jovian side within 
the disk of Europa. 
\textcolor{black}{
With emission from a radially symmetric atmosphere as a reference, 
we searched for an anti-Jovian vs sub-Jovian asymmetry 
with respect to the central meridian 
on the leading side %of the disk, 
and found none.  Likewise, we searched for %symmetrical 
departures from 
a radially symmetric atmospheric emission and found an emission 
surplus centered around 90\grad west longitude, 
for which plausible mechanisms exist. } 
Previous work about the possibility of plumes
on Europa due to tidally-driven shear heating found longitudes with 
strongest local strain rates which might be consistent
with the longitudes of maximum UV emissions.
Alternatively, asymmetries in Europa's UV emission can also 
be caused by inhomogeneous
surface properties, inhomogeneous solar illuminations,  
and/or by Europa's complex 
plasma  interaction with Jupiter's magnetosphere.

\end{abstract}


\section{Introduction}

Jupiter's satellite Europa is an outstandingly interesting body in our
solar system as several lines of evidence point to a subsurface water
ocean under its icy crust. Europa is a differentiated body with an outer 
layer of H$_2$O with a thickness of $\sim$100 km \cite[]{ande07,huss02}.
Its surface is relatively young with
an average age of $\sim$60 Myr and it is 
much less heavily cratered than its two
neighboring satellites Ganymede and Callisto \cite[e.g.,][]{gree04}. 
Evidence for Europa's subsurface ocean comes from observations of geological structures \cite[e.g.,][]{papp99,gree04}. 
Europa's surface is rich in various features,
such as ridges, chaotic regions, and cycloids possibly due to flexing
of its icy crust under tidal forces \cite[e.g.,][]{papp99,nimm02}. 
These features are naturally explained by a subsurface ocean of liquid
water, but it cannot be ruled out that they are alternatively
due to processes in warm, soft ice with only localized or partial 
melting \cite[]{papp99}.
A strong and fully independent argument for a subsurface ocean comes from 
observations of induced magnetic fields \cite[]{khur98,neub98,kive00,zimm00,schi07,saur10} which require a subsurface layer with a sufficiently 
large electrical conductivity. Within Europa's geological and geophysical
context, only saline liquid water is considered a viable candidate to
achieve the required conductivity.

Europa also possesses a thin oxygen atmosphere discovered 
with the Goddard High-Resolution Spectrograph (GHRS) of the Hubble Space
Telescope by \cite{hall95}. 
Observed fluxes of 11.1 $\pm$ 1.7 $\times$ 10$^{-5}$
photons cm$^{-2}$ s$^{-1}$ and 7.1 $\pm$ 1.3 $\times$ 10$^{-5}$
photons cm$^{-2}$ s$^{-1}$ (at the Telescope), corresponding to a 
brightness of 69 $\pm$ 13 and 37 $\pm$ 15 Rayleigh, for the two oxygen 
multiplets \ion{O}{1} 1356  \AA \hspace{1.5mm} and \ion{O}{1} 1304 \AA \hspace{1.5mm}, 
respectively,
implied a very thin molecular oxygen atmosphere with a surface 
pressure of about 10$^{-11}$ bar \cite[]{hall95,hall98}. 
The radiation is excited predominately
by electron impact dissociation of molecular oxygen.
The previous observations were taken on 1994 June 02
at western elongation, i.e. on Europa's trailing hemisphere. 
Subsequent observations 
on 1996 July 21 with GHRS 
on the leading hemisphere  
rendered fluxes of 8.6 $\pm$ 1.2 $\times$ 10$^{-5}$
photons cm$^{-2}$ s$^{-1}$ (54.7 $\pm$ 7.6 R) for \ion{O}{1} 1304 and 
12.9 $\pm$ 1.3 $\times$ 10$^{-5}$ 
photons cm$^{-2}$ s$^{-1}$ (82.1  
$\pm$ 8.3 R) for \ion{O}{1} 1356. Observations on 1996 July 30
confirmed the oxygen atmosphere  on the trailing hemisphere 
with 12.6 $\pm$ 1.3 $\times$ 10$^{-5}$ (80.2 $\pm$ 8.3 R) 
for \ion{O}{1} 1304 and
14.2 $\pm$ 1.5 $\times$ 10$^{-5}$  (90.4 $\pm$9.5 R) for \ion{O}{1} 1356 
\cite[]{hall98}. The joint fluxes 
of both oxygen multiplets were thus 18.2  $\times$ 10$^{-5}$
photons cm$^{-2}$ s$^{-1}$ (106 R) on the trailing side in 1994, 
26.8 $\times$ 10$^{-5}$ photons cm$^{-2}$ s$^{-1}$
(170 R) also on the trailing 
side in 1996, and 
21.5 $\times$ 10$^{-5}$ photons cm$^{-2}$ s$^{-1}$
(137 R) on the leading side in 1996. Based on these observations, 
\cite{hall98} estimated
O$_2$ column densities in the range of 2.4 to 14 $\times$ 10$^{18}$ m$^{-2}$.
Comprehensive reviews on
Europa's atmosphere can be found in \cite{mcgr04,mcgr09}.

Europa's atmosphere is considered to be primarily generated through 
surface sputtering by magnetospheric ions \cite[]{john82,posp89,shi95,shem05}.
Dominant loss processes are atmospheric sputtering and ionization. The balance
of production and loss rates, which both also depend on the
electromagnetic field environment establishes the average atmospheric content.
Estimations of this balance leads to a O$_2$ column density
of $\sim$5 
$\times$ 10$^{18}$ m$^{-2}$ and a neutral O$_2$ loss 
rate of  $\sim$50 kg s$^{-1}$
\cite[]{saur98}. The neutral loss, including H$_2$ from thermal escape,
generates a neutral torus along the orbit of Europa discovered
by \cite{mauk03} and \cite{lagg03}.

Europa is embedded in the Jovian magnetosphere 
whose plasma constantly streams past Europa and its atmosphere. 
The interaction of the magnetospheric plasma with Europa's atmosphere
is primarily responsible for the generation of an ionosphere on 
Europa \cite[]{klio97}. This interaction in addition 
to induction in a possible
sub-surface ocean creates 
magnetic field perturbations
\cite[]{kive97,saur98,schi08} and generates Alfv\'en wings \cite[]{neub99,saur04},
which lead to auroral footprints of Europa in Jupiter's 
upper atmosphere \cite[]{clar02,grod06}. Europa's plasma
interaction modifies not only the electron density but also the 
electron temperature in Europa's atmosphere. Both the 
electron density
and temperature control the \ion{O}{1}1304 \A and \ion{O}{1} 1356  \A
emission rates. Europa's plasma interaction has been 
studied by \cite{saur98,liu01,schi07,schi08}, 
where the latter two models self-consistently include
induction in Europa's possible subsurface ocean.

\begin{figure}
\epsscale{0.60}
%\plotone{../submit_until_April_2010/figure_extern/europa_mcgrath_western.eps}
\plotone{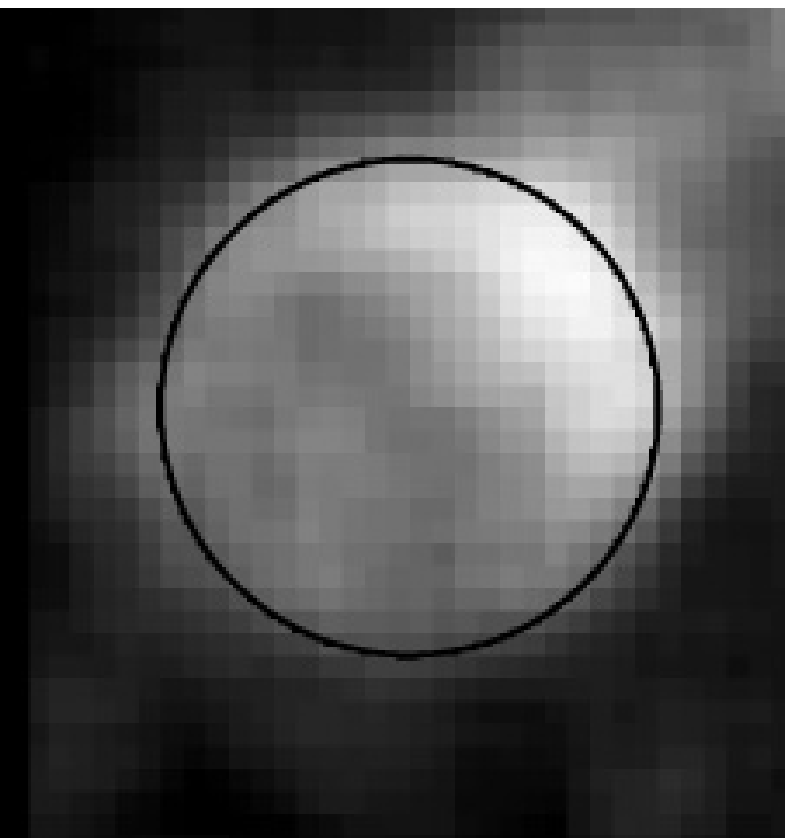}
\caption{Europa's \ion{O}{1} 1356  emission at the trailing hemisphere. 
Image is generated from the superposition of all STIS observations on 1999 October 5 \cite[after][]{mcgr04}. Celestial North is up and Jupiter to the left.   \label{f_western}}
\end{figure}

After the initial spectral observation of Europa's atmosphere by
\cite{hall95,hall98}, the first spatially resolved observations 
of Europa's atmospheric emission
were performed with the Space Telescope Imaging Spectrograph (STIS) 
in 1999 October 5 during five consecutive
HST orbits \cite[]{mcgr04,mcgr09}. A superposition of all exposures
is shown in Figure \ref{f_western}. The emission displays a clear
asymmetry with respect to the sub-Jovian/anti-Jovian side. On the
anti-Jovian side within the disk of Europa occurs a significant enhancement
of the emission. Even though Figure \ref{f_western} suggests time-stationarity
of the emission, the location of the maximum of the emission moves from
Europa's northern hemisphere to its southern hemisphere, but stays
predominantly on the anti-Jovian side \cite[]{mcgr09}. 
This emission is puzzling as it occurs within the
disk of Europa, while for a radially symmetric atmosphere combined
with a radially symmetric plasma interaction, the emission would
be largest near the limb just outside of Europa's disk. 
\cite{cass07} investigate such a stationary surplus of the oxygen emission
and show that a non-uniform distribution of reactive species in 
Europa's porous regolith can result in a non-uniform O$_2$ atmosphere
consistent with Figure \ref{f_western}.

In this work we investigate HST/ACS observations 
of Europa's UV emission taken  on 2008 June 6 
when Europa was at eastern elongation  to further
constrain Europa's atmospheric content and
to investigate if there are also enigmatic inhomogeneities in Europa's UV
emission on the leading hemisphere. In section \ref{s_obs} we
present our observations at eastern elongation and describe how they
were analyzed. In section \ref{s_res} we present the results of our
analysis and discuss their implications. Finally, in section \ref{s_sum}
we summarize our results and point to unresolved issues.

\section{Observations and Analysis}
\label{s_obs}

During HST program 11186, five consecutive orbits of observations with
ACS/SBC prism PR130L on 2008 June 29 were taken (see Table \ref{t_1}). Europa
was at eastern elongation, i.e. the leading side of Europa was visible
from Earth. Each orbit was subdivided into four distinct exposures of
$\sim$10 min. During the time of the observations
Europa also crossed the current sheet of
Jupiter's magnetosphere. Europa's positions with respect to the current
sheet are indicated in Table \ref{t_1}.

Due to the unavailability of STIS during the time of our observations,
we used the prism PR130L \cite[]{bohl00,lars06} to separate reflected 
solar photons from the surface of Europa and the
oxygen multiplets \ion{O}{1} 1356 \AA \hspace{1.5mm} and 
\ion{O}{1} 1304 \AA \hspace{1.5mm} from
Europa's atmosphere.  
The prism covers nominally the wavelength range
larger than 1216 \AA  \hspace{1.5mm} to $\sim$2000 \AA \hspace{1.5mm}
and thus suppresses Ly $\alpha$. Unfortunately, the 
wavelength range larger than 2000 \A is less strongly suppressed
than anticipated leading to a significant 'red leak' \cite[e.g.][]{feld11}.
The sensitivity 
and dispersion properties have been recalibrated  
on-orbit by \cite{lars06}. The dispersion,
i.e. $\Delta \lambda $ per pixel, is a highly non-linear function of 
wavelength. The wavelength $\lambda$ as a function of trace distance $\Delta x$ 
on the detector can be 
written in the form 
\begin{eqnarray}
\lambda(\Delta x)= a_1 + a_2 \;(\Delta x -a_0)^{-1} +a_3 \;(\Delta x-a_0)^{-2} 
+a_4 \;(\Delta x-a_0)^{-3}+a_5 \;(\Delta x-a_0)^{-4}
\end{eqnarray}
with the coefficients
$a_0$ to $a_5$ depending on the object positions $x_{ref},y_{ref}$
in the direct image, i.e.. without prism \cite[]{bohl00,lars06}. 
The spectral trace is additionally shifted in the y-direction on
the detector as a function of wavelength as detailed in \cite{bohl00}
 and \cite{lars06}.

\begin{figure}
\epsscale{1.60}
%\plotone{../../analysis9e/figures/fig_individual_exposures_n08bj.eps}
\plotone{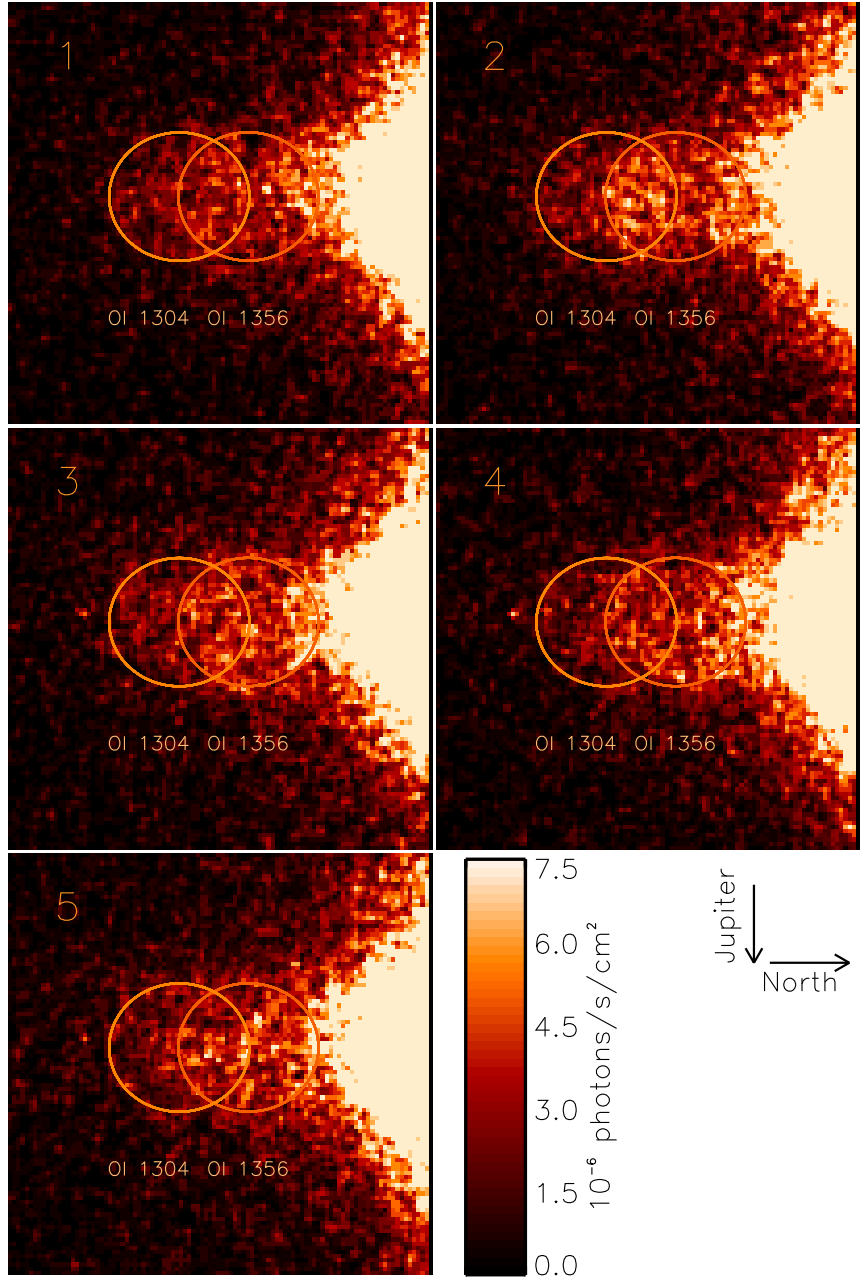}
\caption{Five orbits of HST/ACS/SBC observations on 2008-06-29 (see Table \ref{t_1}). South is to the left and Jupiter is down in each image.   \label{f_obs}}
\end{figure}

In Figure \ref{f_obs} we display the sum of the four exposures
in each orbit separately. 
A superposition of all exposures is shown in the top panel of
Figure \ref{f_model}.
In each image, celestial north is toward the right and defines the
x-direction considered in the present study.
The top of the images is towards the anti-Jovian side of Europa
and is labeled y-direction. The dispersion of the prism 
is in the x-direction, i.e. longer wavelengths lie further
to the right in each image. In each image in Figure \ref{f_obs}
and \ref{f_model},
the two ellipses show the expected locations of Europa's disk for the
\ion{O}{1} 1304 \AA \hspace{1.5mm}and 
\ion{O}{1} 1356  \AA \hspace{1.5mm} emissions
observed with the PR130L prism.
Unfortunately, both images overlap. The overlap is even
somewhat larger than indicated in the overlapping ellipses as the radiation
from Europa's atmosphere off the disk additionally contributes
to the radiation. Therefore we 
analyze the sum of the
\ion{O}{1} 1304 \AA \hspace{1.5mm}and 
\ion{O}{1} 1356  \AA \hspace{1.5mm} emission
in this paper and investigate possible 
asymmetries of the joint emission in
the y-direction only, i.e. the sub-Jovian/anti-Jovian direction, as this
direction is not affected by the overlap.

The sunlight reflected by Europa's surface significantly contributes
to the measured fluxes. In particular, emission longward of
$\sim$2000  \AA \hspace{1.5mm} is not fully filtered out and
leads to the triangle shaped emission on the right hand side of
the image displayed in Figure \ref{f_obs}.
This long wavelength 'red leak' exceeds the fluxes
of the \ion{O}{1} images by roughly two orders of magnitude
much further to the right of the \ion{O}{1} images as
displayed in Figure \ref{f_quantitative_global}.

\begin{figure}
\epsscale{0.9}
%\plotone{../../analysis9e/figures/solar_spectrum.eps}
\plotone{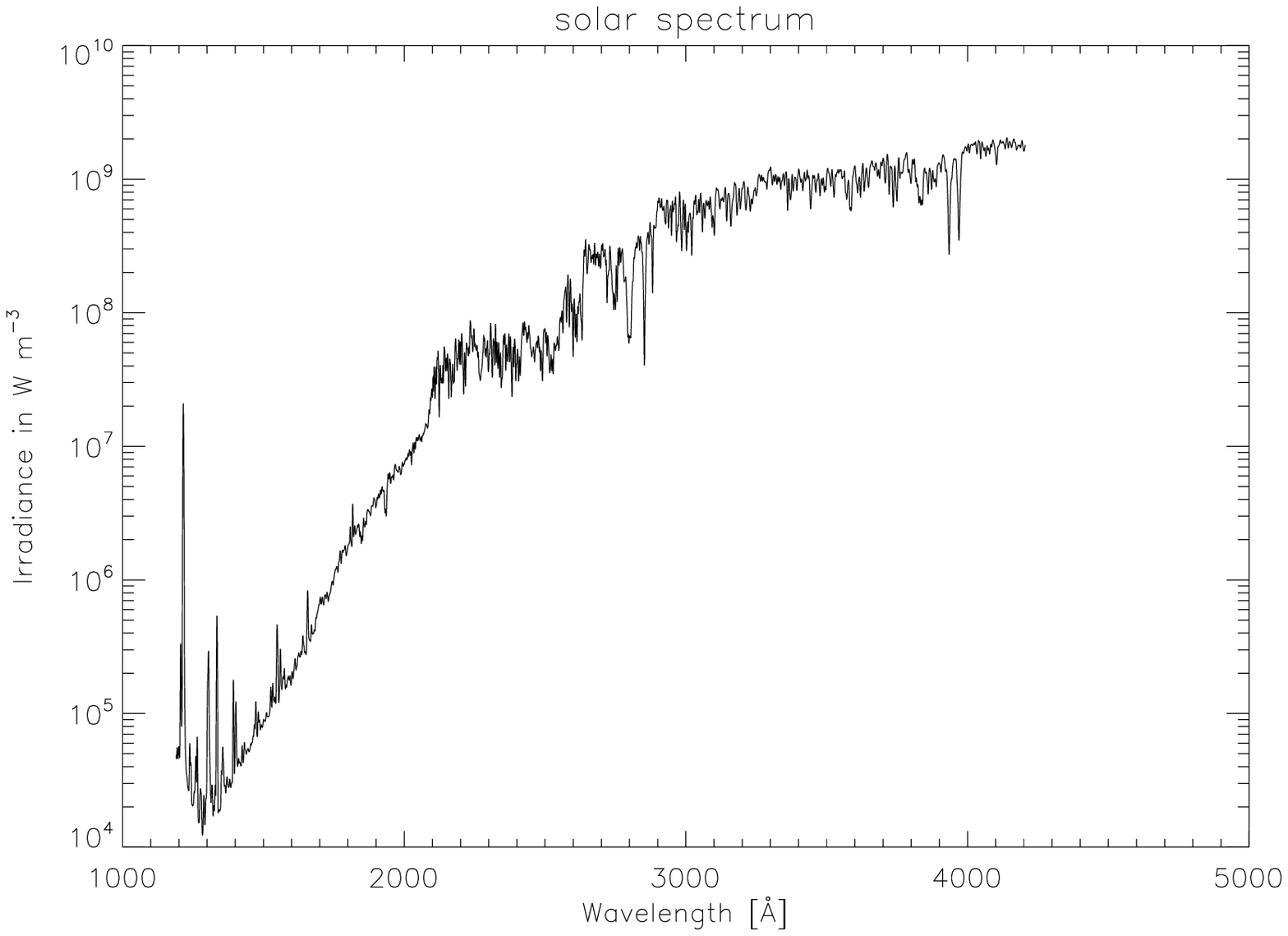}
\caption{Solar spectrum after \cite{wood96}   \label{f_solar_spectrum}}
\end{figure}

\begin{figure}
\epsscale{1.80}
%\plotone{../../analysis9e/figures/fig_allexposures_and_model_n08bj.eps}
\plotone{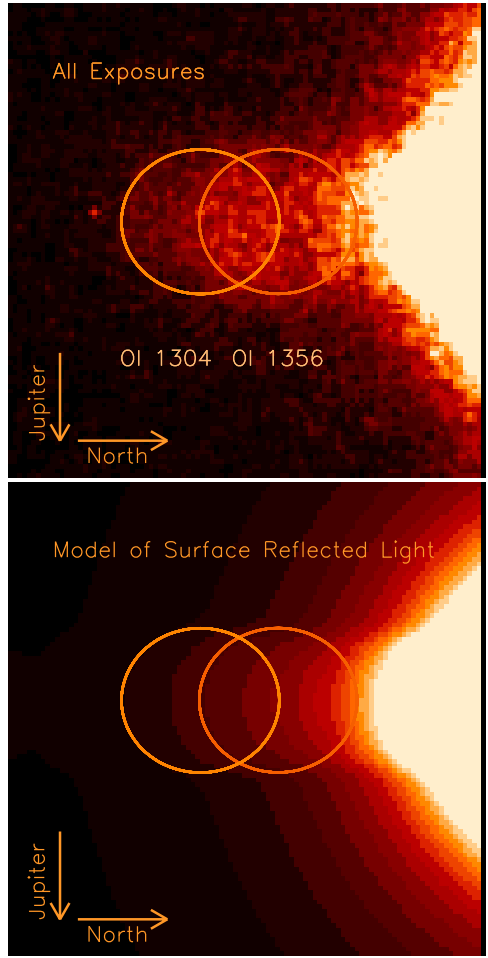}
\vspace*{-7cm}
\caption{Sum of all exposures (top) and modelled reflected solar light 
from disk of Europa (bottom).  The same color code is applied
as in Figure \ref{f_obs}. North is to the right 
and Jupiter is down in each image.  
\textcolor{black}{
The color scale is identical to 
the color scale of Figure \ref{f_obs}}.
\label{f_model}}
\end{figure}

\begin{figure}
\epsscale{1.0}
%\plotone{../../analysis9e/figures/fig_allover_quantitative_n08bj.eps}
\plotone{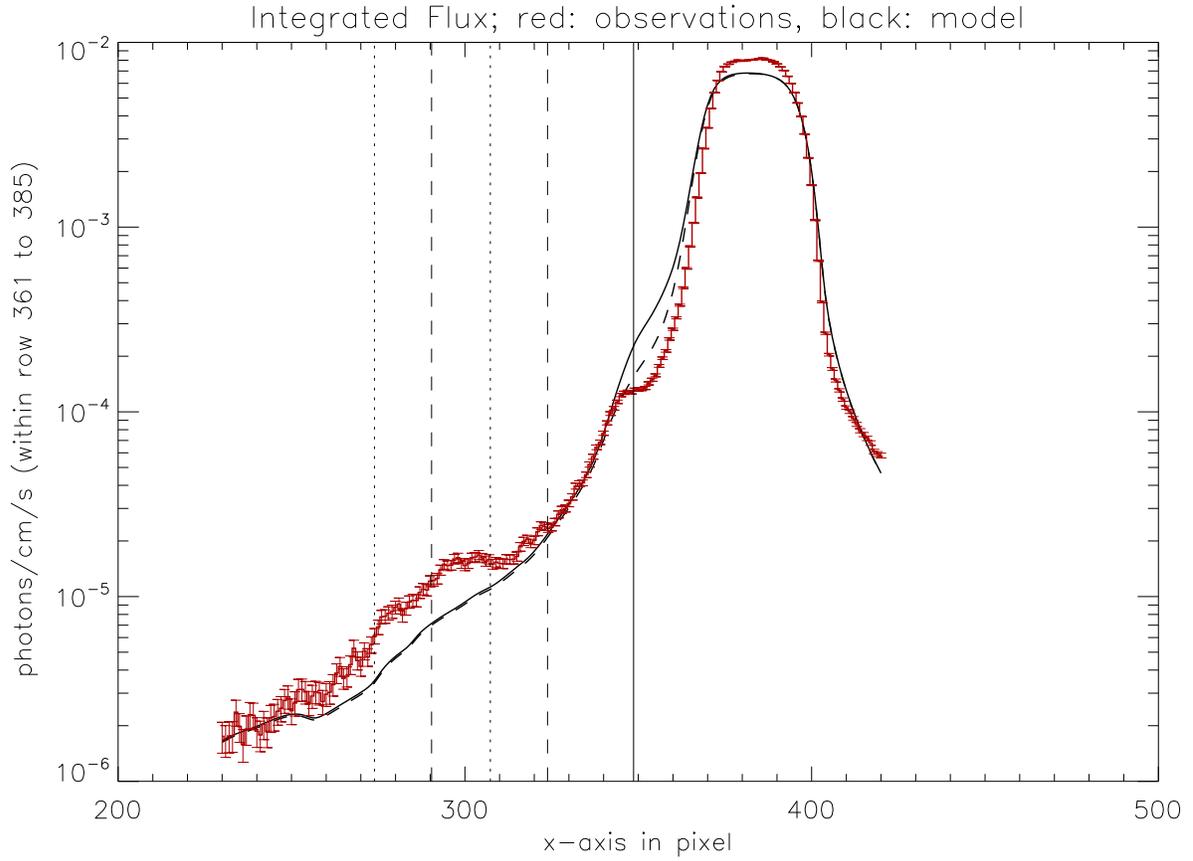}
\vspace*{0cm}
\caption{Observed fluxes (red) along trace (x-axis)
summed along y-axis calculated with a superposition of all exposures. 
Modelled fluxes of solar light reflected from Europa's surface 
is shown as black solid line. 
\textcolor{black}{
Reflected light calculated with an alternative model
is displayed as black dashed line (further explanations in text).}
Vertical dotted lines show area of O1304 image and vertical  dashed lines displays area of \ion{O}{1} 1356  image. \label{f_quantitative_global}}
\end{figure}

In order to account for the contribution of the reflected sunlight
to the \ion{O}{1} images, we model its photon flux $F_S$ as a function
of position  $(x,y)$ on the detector.
The reflected
and dispersed solar light can be calculated as
\begin{eqnarray}
F_s(x,y)=\int_{\mbox{solar spect}} d\lambda \; p(\lambda) t(\lambda) 
a(\lambda) D[x 
- (\Delta x (\lambda) + x_{E}),y - (y_d(\Delta x(\lambda))+y_{E})]
\label{e_model}
\end{eqnarray}
with the dispersion $d \lambda/ d (\Delta x) $ \cite[]{lars06} and 
$\Delta x(\lambda)$ 
being the trace distance of the  
dispersed photon flux in x-direction as a function of wavelength $\lambda$
compared to the location of the undispersed flux without prism. 
The location of the spectral trace is also shifted in the y-direction by
$y_d(\Delta x)$ \cite[]{lars06}. The coordinates $x_{E}$, and $y_{E}$ refer to
the center of Europa's disk without dispersion.
$D[x(\lambda),y(\lambda)]$ represents the disk of Europa
with $D[x,y]=1$ within the
disk, and $D[x,y]=0$ outside of the disk.  Since we used a prism, $D$ is
a function of wavelength $\lambda$. The solar photon flux $p(\lambda)$ is 
given by \cite{wood96} \textcolor{black}{
and displayed in Figure \ref{f_solar_spectrum}},
$t(\lambda)$ is the throughput, and $a(\lambda)$ Europa's albedo.
The throughput for $\lambda < 2000$ \A is provided by \cite{lars06}
and constrained for $\lambda > 2000$ \A based on observations of 16 CygB
as detailed in Appendix \ref{ss_throughput}. 
The flux $F_s(x,y)$ is additionally convolved with a point spread function
as described in Appendix \ref{ss_psf}.

Since no direct image of Europa was taken during the
HST campaign 11186, the location of the direct image has to be
reconstructed from the dispersed images. For this purpose the strong
red leak is very useful. The long wavelengths reflected solar light 
generates in the obtained observations an easily discernible region 
of maximum flux corresponding to the size of Europa 
\textcolor{black}{when plotted with a linear color scale, 
owing to the very low dispersion of the prism at red leak wavelengths}. 
For this
long wavelength image of Europa, we determine its 
coordinates and then use expression (\ref{e_model}) to
subsequently determine $x_E$ and $y_E$. The latter step is possible
despite the uncertainties in the throughput for long wavelengths
since the dispersion is very large at long wavelengths, e.g. 
$\sim250$ \A pxl$^{-1}$ at 4000 \AA. Therefore uncertainties in the slope
of the throughput at long wavelengths have very small effects on
the position of the dispersed long wavelength image of Europa.

Europa's albedo as a function of wavelength is only partially constrained
by previous observations. \cite{hall98} derive \textcolor{black}{a disk
averaged albedo of 0.015 $\pm$ 0.004 on the leading side  
and similar albedo values on the trailing side
at $\lambda=1335$ \A. }
Based on an analysis of STIS observations of Europa's
trailing side in 1999 (\textcolor{black}{ID 8224, PI M. McGrath}), 
the albedo stays remarkably flat in the FUV between 1400 and 1700 \A at
values near 0.016. 
The constancy of the albedo values
in this wavelength range \textcolor{black}{(not shown here) 
is derived similarly to the analysis of STIS 
observations of Ganymede (ID 7939, PI W. Moos, \cite{feld00a}). }
\cite{hend98} find that the 
albedo on the leading side at central meridian of 120$^{\circ}$ W increases
approximately linearly from $\sim$0.2 to $\sim$0.65 
within $\lambda = 2200$ \A to $\lambda = 3200$ \AA.
\cite{noll95} and \cite{hend98} find a lower albedo
on the trailing side which changes from $\sim$0.08 to $\sim$0.17 
within $\lambda = 2200$ \A to $\lambda = 3200$ \AA.
The general trend that the albedo on the leading side is larger compared
to the trailing side is also seen by \cite{nels87}. The authors
observe albedo variations by roughly a factor of two from the leading
to the trailing hemisphere within $\lambda =$ 2400 \A to 3200 \AA. The maximum
values of the albedo on the leading side are 
0.18 (2400--2700 \AA), 0.24 (2800--2900 \AA) 
and 0.32 (3000--3200 \AA).  
\textcolor{black}{ \cite{hans05} derive an albedo of $\sim$1\% 
(without error bars)
near 1335 \A and at 94$^{\circ}$
phase angle.}

\begin{figure}
\epsscale{0.7}
%\plotone{../../analysis9e/figures/albedo_n08bj.eps}
\plotone{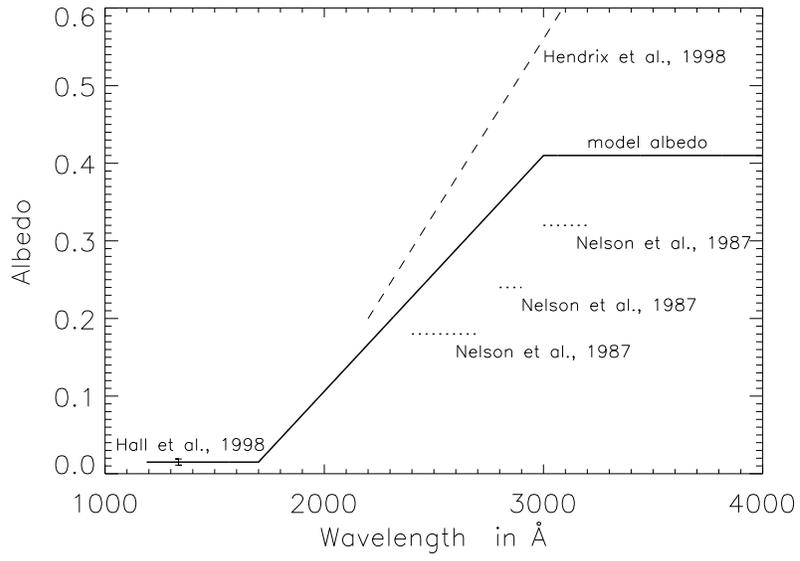}
\vspace*{0cm}
\caption{Model albedo for Europa's leading side as used in our analysis. \label{f_albedo}}
\end{figure}

Our analysis pertains to the leading side, 
i.e. for the central meridian 90$^{\circ}$, where we 
assume the model albedo 
shown in Figure \ref{f_albedo}.  
\textcolor{black}{
In the absence of albedo measurements
within 1700 \A and 2200 \AA, we choose to construct an albedo model
which is as simple
as possible and is still consistent with the constraints
described in the previous paragraph.
For $\lambda < 1700$ \A our model albedo is constant at a value of 0.015. 
For 1700 \A $< \lambda < $3000 \A the albedo rises linearly 
from 0.15 to 0.41 and
stays flat for $\lambda > $3000 \A (see Figure \ref{f_albedo}). 
Further details on how the quantitative values of the linear rise of
our model albedo are chosen, will be discussed in
subsequent paragraphs where we compare the observed fluxes with the
modeled solar reflected light.
}

With the procedure and model inputs described in the previous paragraphs,
we calculate how the 
solar light reflected from the surface is dispersed with the prism PR130L
and appears on the detector.
The resultant flux is
shown in Figure \ref{f_model}, lower panel.
For comparison we show the observed fluxes
of all exposures superposed in Figure \ref{f_model}, top panel.
Several features of the modeled reflected light can be seen
in the observations, such as the bright emission 'triangle' to the right 
of the \ion{O}{1} emissions due to reflected long wavelengths light. The extension
of the emission in y-direction within the triangle is due to the
scattering within the prism and described by the point spread function
constrained in Appendix \ref{ss_psf}. It is also clearly visible
that there is a contribution from the
reflected light also within the \ion{O}{1} images.

In Figure \ref{f_quantitative_global} we quantitatively show the modeled
reflected surface flux and the observed flux of all exposures
superposed along the direction of the dispersion.
To reduce statistical fluctuations,
we integrated the fluxes in the y-direction. 
\textcolor{black}{To minimize the contribution from the off-disk emission,
the integration
was performed over 25 pixels from 
rows 361 to 385 (i.e., somewhat
less than the diameter of Europa on the detector, which is 
approx. 30 pixels in the y-direction). 
}
The integrated modeled (black) and observed fluxes (red)
are plotted as a function of trace distance (i.e., in the x-direction) in
Figure \ref{f_quantitative_global}. The two vertical dashed lines show the
locations of the \ion{O}{1} 1356 \A image and the two vertical dotted lines the
locations of the \ion{O}{1} 1304 \A image corresponding to the size
of Europa's disk. 
The solid vertical line in Figure 
\ref{f_quantitative_global} shows the 2000 \A 'influence line'. Reflected
light from the disk of Europa with wavelengths larger than 2000 \AA, i.e.,
from the red leak, would lie strictly to the right of this line 
if scattering were not to take place.

\textcolor{black}{
We use the observed fluxes displayed in Figure
\ref{f_quantitative_global} 
to constrain our model albedo (see Figure \ref{f_albedo}). 
We assume a simple linear increase of the model albedo for wavelengths larger
than 1700 \AA. The quantitative values of this increase has been chosen
to match the observed fluxes 
sufficiently far away from the wavelength ranges which correspond to the
\ion{O}{1} emission from Europa's atmosphere (i.e. outside of the
vertical dotted and dashed
lines). The modeled
reflected light is shown as a solid black line in Figure 
\ref{f_quantitative_global}, which fits the observations outside
of the atmospheric \ion{O}{1} wavelengths ranges in general well. 
However, near the solid vertical line the surface reflected
light overestimates the observed fluxes and does not reproduce
the 'knee' in the observations.
A principal reason could be a dip in the albedo around 
2000 \AA. But any uncertainty in
the model albedo $a$ can equally well be reflected in
an equal uncertainty in the throughput $t$ 
when working with the prism observations.  
Only the product of $a$ and $t$ constrains
the observed fluxes (see equation (\ref{e_model})). 
The throughput at wavelengths around and larger than 2000 \A is also not
well constrained (see discussion in Appendix \ref{ss_throughput}).
To improve the fit of the modeled reflected light we introduce a
throughput dip of a factor 1/3 within 1800 \A and 2300 \A (shown
as dashed blue line in Figure \ref{f_throughput}). This modified
throughput leads to a modified spectrum of reflected and dispersed
light from the surface of Europa and is displayed as the dashed
black line in Figure \ref{f_quantitative_global}. The spectrum
with the throughput dip reproduces the knee in the observed
spectrum much better. We will call it in the remainder of this
paper 'alternative' model of reflected surface light. It will 
provide in subsequent parts of the paper a rough estimate for the
consequences of the uncertainties in the model albedo and/or throughput,
i.e., it provides an estimate for systematic errors introduced in the
subtraction of the reflected light.
}

At some distance away from the locations
of the oxygen images (constrained by the vertical dashed and dotted
lines) the observation are close to our modeled reflected light.
The surplus of the observed emission compared to the model solar reflected
light within and near the dotted and dashed lines
is the \ion{O}{1} 1356  \A 
and    \ion{O}{1} 1304 \A radiation emitted from Europa's atmosphere. 
It should be noted that in Figure \ref{f_quantitative_global} the 
flux scale is logarithmic,
thus a surplus at x=280 appears
roughly ten times larger than the same surplus at x=330.

\textcolor{black}{
We note that any possible contribution by the geocorona plays a negligible
role in our analysis. The \ion{O}{1} 1356 \A of Earth's atmosphere is below the
HST orbit. As we observe Europa near Jupiter opposition, i.e.,
during HST orbit night, the \ion{O}{1} 1304 \A emission only contributes
1 or 2 Rayleighs uniformly on the detector.}

\begin{figure}
\epsscale{1.6}
%\plotone{../../analysis9e/figures/fig_individual_quantitative_n08bj.eps}
\plotone{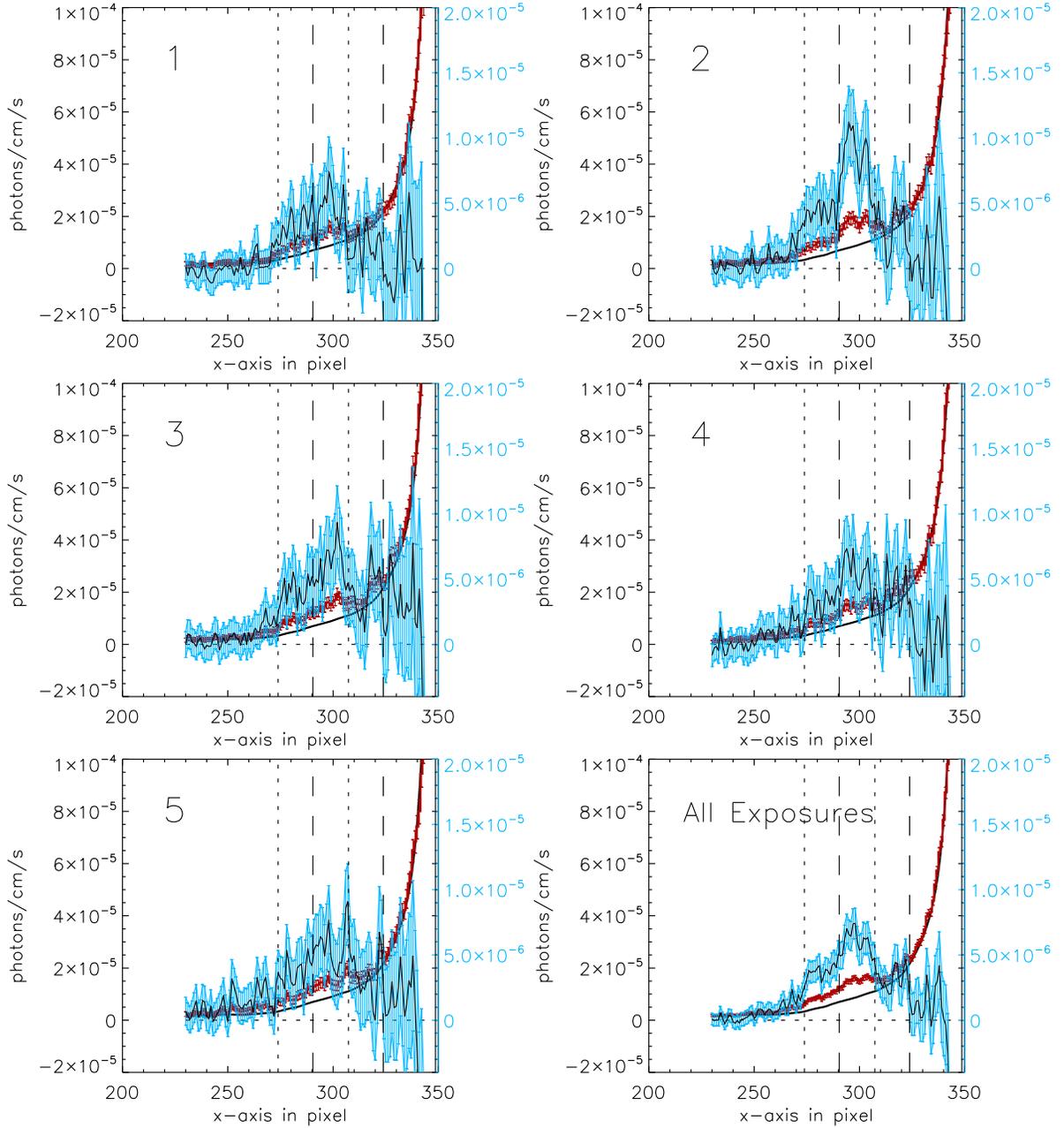}
\vspace*{-2cm}
\caption{Quantitative comparison of observed fluxes (red) along trace (x-axis)
summed along  y-axis and modelled solar reflected light (black). The difference
between the observed and reflected light is the emission from Europa's
atmosphere shown 
\textcolor{black}{
as black thin line with error bars in blue.}
 The emission from Europa's atmosphere is multiplied
by a factor of five for better visibility (the corresponding values
are displayed on the right hand side of each plot, respectively). 
Vertical dotted lines show area of \ion{O}{1} 1304 \A image and vertical  dashed lines displays area of \ion{O}{1} 1356  \A image. \label{f_quantitative_individual}}
\end{figure}

In the next step we display in Figure \ref{f_quantitative_individual}
the observed fluxes for each HST orbit separately.
For comparison we also show the modeled solar reflected fluxes.
The fluxes are integrated
between rows 361 to 385 similar to Figure \ref{f_quantitative_global}. 
All error bars include statistical errors only. The signal
to noise ratio is calculated by 
\textcolor{black}{
$S/N=(S-B-D)/\sqrt{S+B+D}$
}
with S and B corresponding
to the total counts of the signal and the background, respectively,
and D to the total counts of the ACS/SBC dark noise (S, B and D are summed 
over exposure time and pixel under consideration, respectively). 
The dark noise count rate is
1.0 $\times 10^{-5}$ cts$^{-1}$ s$^{-1}$ pixel$^{-1}$
\cite[]{cox04}.
A more detailed representation of the emission from Europa's
atmosphere is given by the integrated fluxes on a linear scale
spanning wavelengths below 2000 \AA.
The solid vertical line in Figure 
\ref{f_quantitative_individual} shows the 2000 \A 'influence line'. 
The two vertical dashed lines show the
locations of the \ion{O}{1} 1356 \A image and the two vertical dotted lines the
locations of the \ion{O}{1} 1304 \A image corresponding to the size
of Europa's disk.

The differences between the modeled surface reflected light (black) and
the observed fluxes (red) in Figure \ref{f_quantitative_individual} are due to
the emission from Europa's oxygen atmosphere (blue).
We multiplied the values of emission from Europa's atmosphere by a factor
of five to more clearly display it within the individual plots.
The corresponding values for the atmospheric emission are shown
on the right hand side of each panel in blue.
The model fluxes (black) to the right of the \ion{O}{1} emission region and to the left 
of the 2000 \A influence line match 
all individual exposures and the superposed exposure (red) 
fairly well and thus
demonstrate that the solar reflected light contributions
does not change from orbit to orbit. 
In contrast, the observed emissions in blue 
display apparent time-dependence. The observed fluxes
change as a function of orbit number as will be analysed in
more detail in the following section.  
The observations also show that there is excess emission beyond 
the disk of Europa due to the spatial extension of Europa's atmosphere
\textcolor{black}{as visible to the left of the \ion{O}{1} 1304 \A disk, i.e. 
to the left of the left dotted vertical line. Any excess emission
to the right of the \ion{O}{1} 1356 \A disk, i.e. to right of the
right vertical dashed line, might be buried within the large error
bars in this region.}
The signal to noise ratio
of the radiation from Europa's atmosphere displayed in Figure
\ref{f_quantitative_individual} is however partially as low as two
as apparent in  the large error bars. Therefore we will investigate
in the remainder of this work areas corresponding to a larger 
number of pixels for a  better signal to noise ratio.

\section{Results}
\label{s_res}

\subsection{Radial/equatorial structure of atmosphere}

\begin{figure}
\epsscale{1.0}
%\plotone{../../analysis9e/figures/fig_limb_n08bj.eps}
\plotone{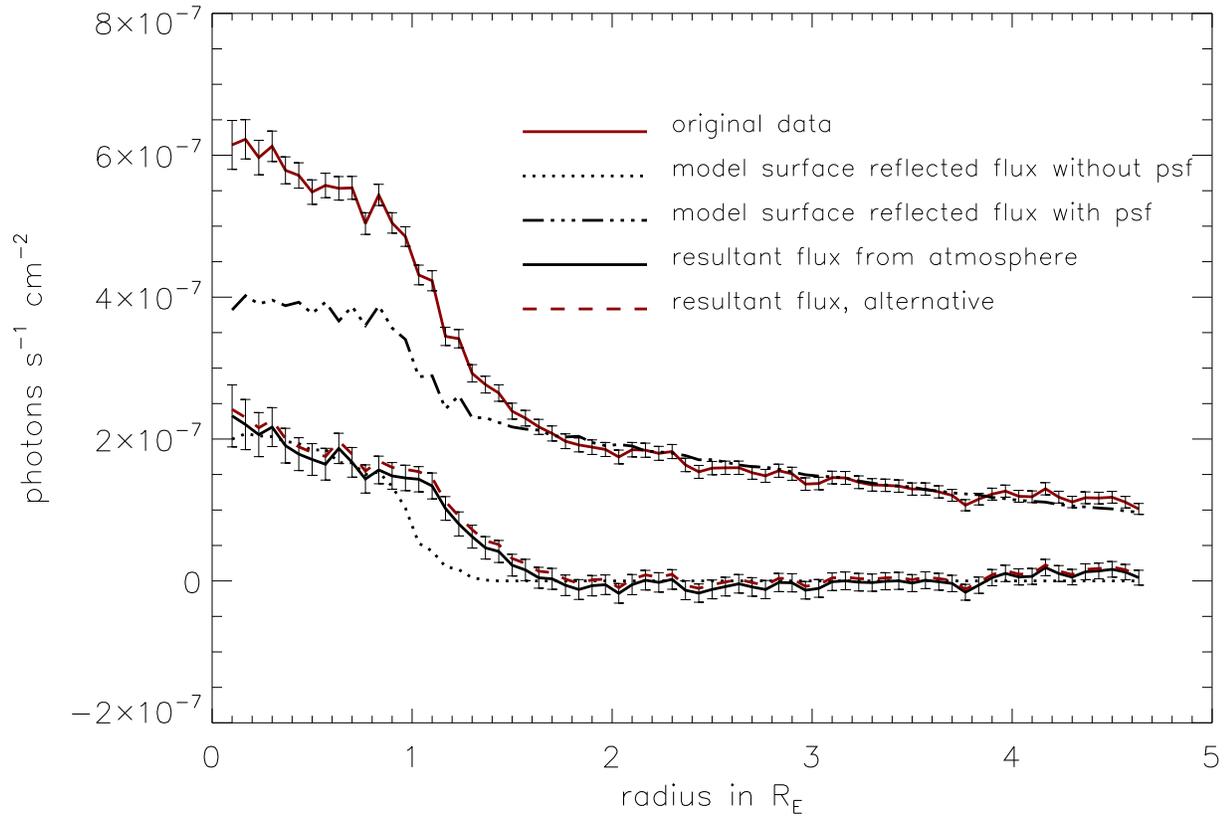}
\vspace*{0cm}
\caption{Average emission of Europa's atmosphere as a function of
radial distance from Europa's center within columns 272 to 326
have been considered.\label{f_limb}}s
\end{figure}

In Figure \ref{f_limb} we show the average emission as a function of radial
distance from the center of the disk of Europa, which we calculate
as average emission within thin 
radial shells of both disks (\ion{O}{1} 1304 \A and 1356 \AA).
\textcolor{black}{The emission is normalized 
to the number of pixels in each shell.
The radiation within the shells is calculated as the difference
in radiation
within radius $r + dr$ and of radius $r$. For radial values where 
the two 
\ion{O}{1} 1304 \A and 1356 \A areas overlap, 
we do not calculate the overlapping
part twice. No attempt is made to separate the \ion{O}{1} 1304 
emissions from the \ion{O}{1} 1356 emissions 
in the overlapping regions.  Instead our model instrument 
response replicates the overlap as part of the fitting process.}
We restrict ourselves to pixels within columns 
274 to 323
%323 depending on how to round 
in case the shells extend beyond these columns. These columns
\textcolor{black}{just embrace the \ion{O}{1} 1304 \A disk to the left and 
the \ion{O}{1} 1356 \A disk to the right in x-direction}.
If we were to integrate further 
\textcolor{black}{to the right in the x-direction}
(direction of the dispersion) we would also integrate
pixels 
whose emission is significantly contaminated by 
%other sources than \ion{O}{1} lines.
%We would then particularly include pixels where 
the red leak. %plays
%an important role, which would strongly contribute in Figure \ref{f_limb}.
\textcolor{black}{To avoid over-counting the \ion{O}{1} 1304 \A 
compared to the \ion{O}{1} 1356 \A emission, we do not integrate 
further to the left of the \ion{O}{1} 1304 \A image.}
Consequently, the profiles in Figure \ref{f_limb}
is more constrained by equatorial than by polar off-disk
emission. 

The red line in Figure \ref{f_limb} shows the
original observations. The dotted line shows the reflected surface
light without inclusion of the point spread function.  
The apparent model emission extends to
radial distances larger than 1 R$_E$
even though the surface reflected light is generated within 1 R$_E$.
The reason is that the dispersion of the
continuous solar spectrum also generates
emission out side of the location of the \ion{O}{1} images.
The solid black line shows the model surface reflected light including
the point spread function. It shows that the scattering of
photons within the prism is a significant effect as can be
seen in Figure \ref{f_model} as well. The model reflected light fits
the observations at large radial distances remarkably well
and thus independently confirms the quality of the constructed
point spread function (see Appendix). The difference between
the solid red and the black line is the excess emission from 
Europa's atmosphere (shown as dashed solid line).  
It demonstrates non-negligible emission above
the disk of Europa which tapers off towards  
$\sim$1.5 - 1.7$R_J$. 
\textcolor{black}{In Figure \ref{f_limb} we also show the resultant flux
from Europa's atmosphere calculated with the alternative throughput
model as red dashed lines. These fluxes differ by only $\sim$10\% 
or less from the fluxes derived with our standard model.
}

\subsection{Atmospheric emission, individual orbits}
\begin{figure}
\epsscale{1.6}
%\plotone{../../analysis9e/figures/fig_individual_exposures_rm_n08bj.eps}
\plotone{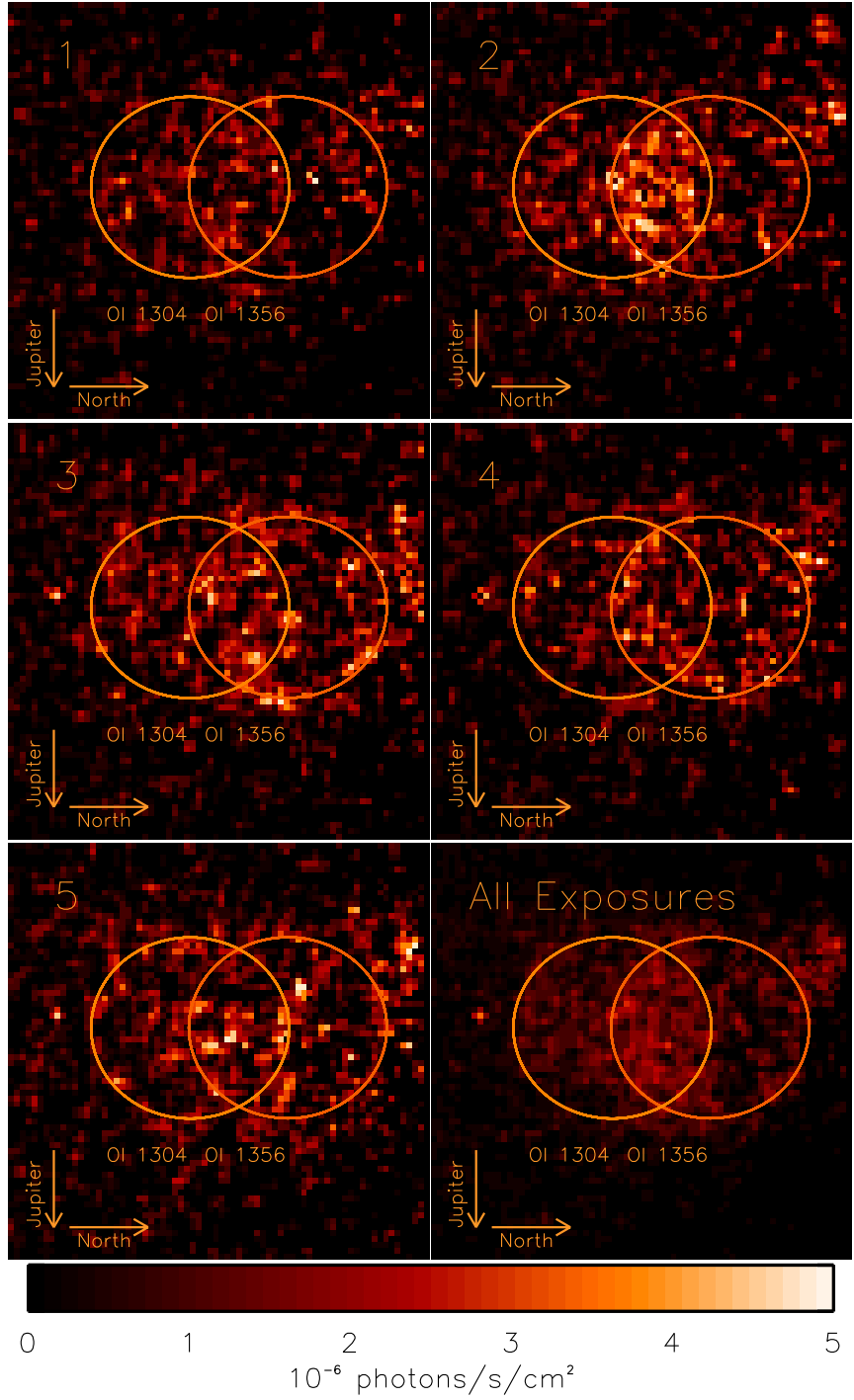}
\vspace*{0cm}
\caption{Individual exposures after subtraction of reflected solar 
light on disk of Europa.  North is to the right and Jupiter is down in each image.  
%The color scale 
%is similar to
%Figure \ref{f_obs}, but enhanced by a factor of 1.5
%to more clearly display the fluxes.
\label{f_rm}}
\end{figure}

The \ion{O}{1} 1304 \A and the \ion{O}{1} 1356 \A
emission from Europa's atmosphere can be reconstructed
by subtraction of the reflected and dispersed solar light
of the surface from the observations (see Section \ref{s_obs}).
The emissions from Europa's atmosphere for all
five orbits and the sum of all exposures are shown
in Figure \ref{f_rm}. The color scale is similar to
Figure \ref{f_obs}, but enhanced by a factor of 1.5
to more clearly display the flux variations.
We note that individual pixels
or small cluster of pixels should be interpreted
with much caution due to their low signal to noise ratio.
The general impression from the images in Figure \ref{f_rm} 
is that most of the flux occurs within or close to 
both \ion{O}{1} images. The images also show time dependence
with orbit 2 and 3 having maximum fluxes. The emission
emerges mostly within the disk of Europa, but it also extends
beyond the disk as \textcolor{black}{
due to a finite atmospheric scale height reported in} 
previous observations \cite[]{hall95,mcgr04} and modeling \cite[]{saur98}.

\subsection{Total fluxes}
\label{ss_totalfluxes}
\begin{figure}
\epsscale{1.5}
\hspace*{-2cm}
%\plotone{../../analysis9e/figures/fig_currentsheet_n08bj.eps}
\plotone{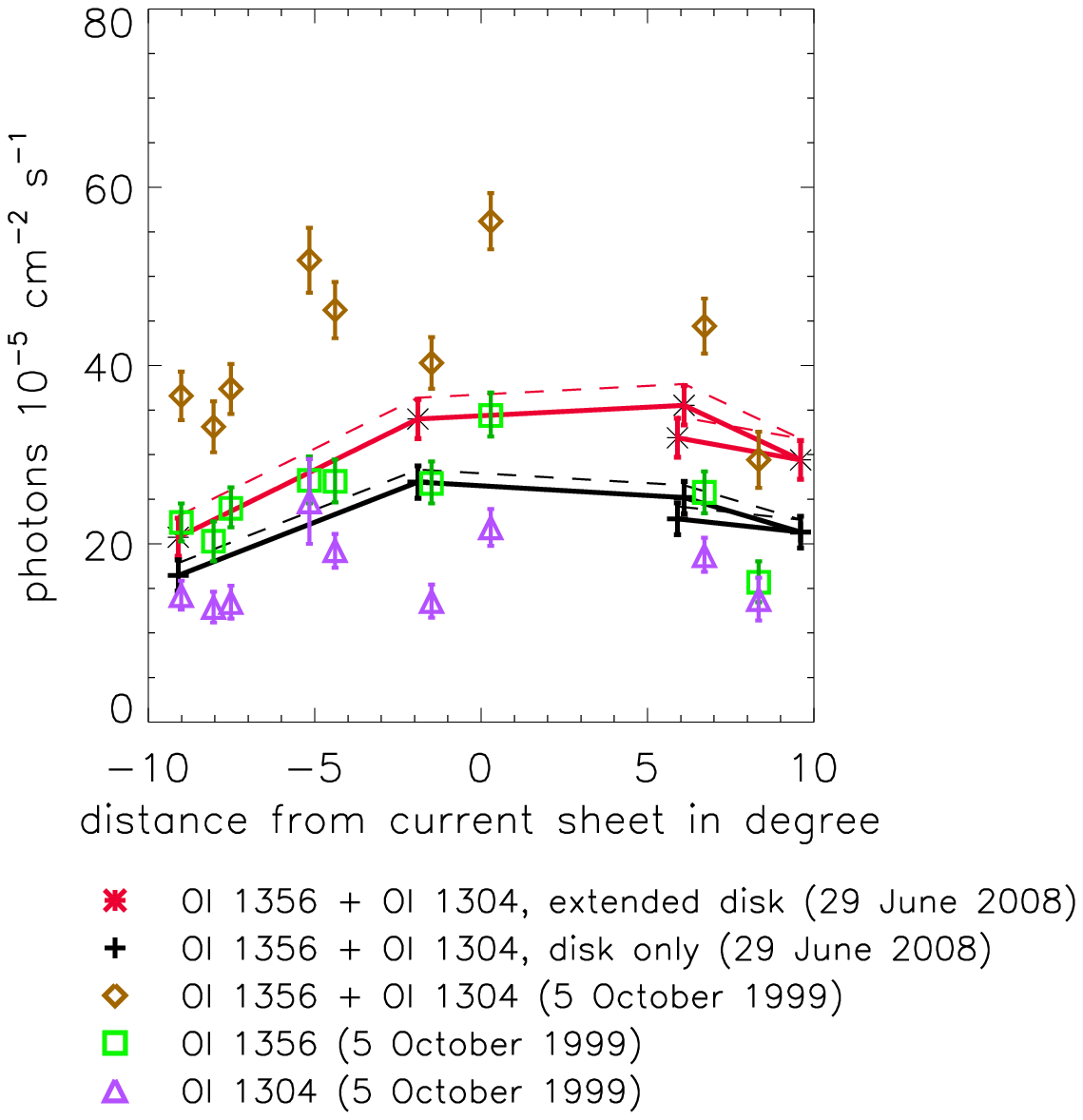}
\vspace*{0cm}
\caption{Total fluxes as a function of nominal distance with respect to 
Jupiter's magnetic dipole equator. 
Flux on disk and limb for 2008 June 29  observations ($r<1.3  R_{Europa}$).  
\textcolor{black}{
The dashed lines represent the total fluxes when the alternative 
throughput model is used (see text).
}
\label{f_currentsheet}}
\end{figure}

Due to the low signal to noise ratio for
individual pixels or small groups of pixels, it is necessary
to integrate the emissions over larger areas.
Therefore we calculate the
total flux from Europa's atmosphere for each orbit
within an extended disk
including a radial distance of
1.3 $R_E$ from Europa's center with Europa's radius $R_E$ = 1569 km.
In Figure \ref{f_currentsheet} the total fluxes 
are shown in red as a function of Europa's position
with respect to the current sheet of Jupiter's magnetosphere.
Jupiter's magnetic moment is tilted with respect to Jupiter's spin
axis by $\sim$9\grad . Jupiter's fast rotation period of 9 h and 55 min
is responsible for large centrifugal forces, which confine the magnetospheric
plasma into a disk around Jupiter's centrifugal equator, called
'current sheet' or 'plasma sheet'. The current sheet is 
close to the magnetic equator while Europa's orbital plane is within
Jupiter's spin equator. Thus the current sheet rocks above and below
Europa within 11.23 h (the synodic rotation period of Jupiter as seen
from Europa). Therefore Europa is exposed to different plasma
conditions, in particular varying electron
densities, within the 11.23 h. This effect is clearly visible in 
Figure \ref{f_currentsheet}. The maximum emission of $\sim$36 $\pm$ 2.2 
$\times$ 10$^{-5}$ photons cm$^{-2}$ s$^{-1}$ 
is reached when Europa is in the center of
the current sheet and minimum values of $\sim$21 $\pm$ 2.1 $\times$ 10$^{-5}$
photons cm$^{-2}$ s$^{-1}$ occur when Europa is at the edge of the current sheet.
These fluxes correspond to $\sim$230 $\pm$ 14 and $\sim$130 $\pm$ 14 Rayleigh,
respectively (see also Table 1). 
\textcolor{black}{
The quoted error bars contain statistical
errors only. In Figure \ref{f_currentsheet}, we also show the fluxes
calculated with the hypothetical dip in the throughput curve as alternative
model with dashed lines. The alternative fluxes 
differ from the standard model by 
as much as the width of the statistical errors. The difference between the
standard and the alternative model can be considered as a rough measure
for the systematic error introduced by subtracting the surface reflected
light, which contains uncertainties due to the partially unconstrained
albedo and red leak.
}

In black we show the fluxes within the disk of Europa, i.e.,
within 1 $R_E$. The associated values are roughly 3/4 of the fluxes
of the extended disk. We note, as both \ion{O}{1} images overlap, some of the off disk
emission from the real atmosphere falls within the \ion{O}{1} disks of the
two dispersed \ion{O}{1} images. Thus the quoted ratio is an upper limit for
the real disk emission.

We can compare the current sheet dependence of the total fluxes to 
the observations from 1999 October 5 taken with STIS on the
trailing side \cite[]{mcgr04,mcgr09}. Similarly to the ACS data,
we also integrate the STIS data within 1.3 $R_E$.
The total fluxes from the 1999 October 5 campaign are shown as brown
diamonds in Figure \ref{f_currentsheet}. The green squares in Figure
\ref{f_currentsheet} show the 
current sheet dependence of the \ion{O}{1} 1356 \A emission and the violet
triangles those of the \ion{O}{1} 1304 \A emission. The ratio of the latter two
is roughly two to one.
The 1999 observations
also show a clear dependence on Europa's position with respect
to the current sheet 
\textcolor{black}{for the \ion{O}{1} 1356  \A emission, while  
the dependence of the \ion{O}{1} 1304 \A emission is ambiguous}. Their total fluxes 
are roughly 30\% to 40\%  larger
than the values derived from the  2008 ACS observations of this work.
The combined \ion{O}{1} 1356  and \ion{O}{1} 1304 fluxes 
of 137 Rayleigh observed by \cite{hall98} with GHRS on the leading
side are consistent with the 2008 ACS observations outside of the current sheet
and $\sim$50\%  smaller than the 2008 ACS values when Europa is in the center of
the current sheet.

We can use the observed fluxes within the range of 170 to 270 Rayleigh
and estimate average column densities. Assuming for simplicity a constant
electron density of 40 cm$^{-3}$ and temperature of 20 eV similar
to \cite{hall95} and  neglecting the electrodynamic interaction,
we can derive a lower limit for the O$_2$ column
density of 6 to 10 $\times$ 10$^{18}$ m$^{-2}$.  
\textcolor{black}{Our derived column densities show somewhat less variations,
but are still roughly consistent compared to
the column densities   of 
3.5 to 11 $\times$ 10$^{18}$ m$^{-2}$ derived by \cite{hall98}.
Note that the observed electron density
upstream of Europa varies between 18 to 250 cm$^{-3}$ \cite[]{kive04}. }

\subsection{Search for asymmetries on leading  hemisphere}
\begin{figure}
\epsscale{1.0}
%\plotone{../../analysis9e/figures/fig_asym_n08bj.eps}
\plotone{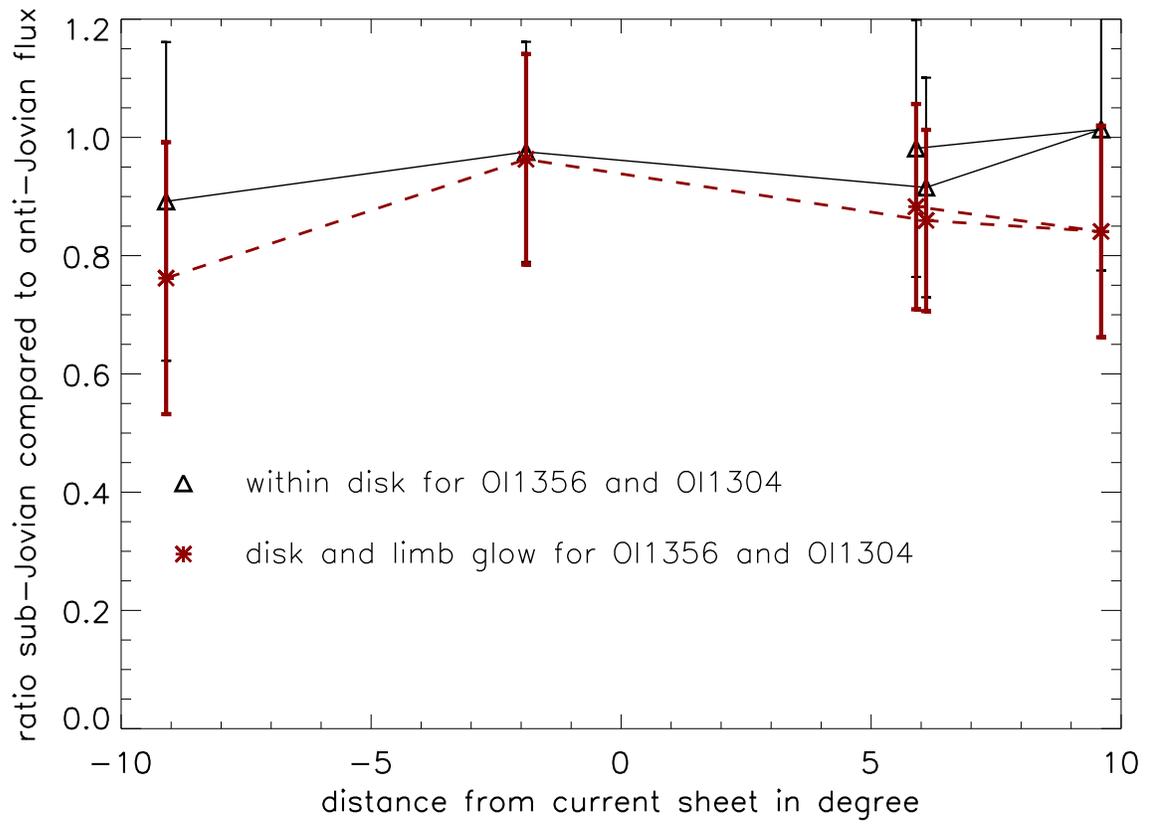}
\vspace*{0cm}
\caption{Ratio of sub-Jovian flux compared to anti-Jovian flux
for 2008 June 29 observations. For the fluxes
within the disk no asymmetry is visible.  \label{f_asym}}
\end{figure}

The spatially resolved observations of Europa's atmosphere on
the orbital trailing hemisphere in 1999 show a pronounced surplus
of emission on the anti-Jupiter side (see Figure \ref{f_western}).
We quantitatively searched for similar asymmetries on Europa's 
leading hemisphere. Fortunately, the dispersion is in the north-south direction
and we can thus well analyze possible asymmetries with respect
to the sub-Jovian and anti-Jovian side. In Figure \ref{f_asym}, we
show the ratio of the total flux within the sub-Jovian side 
compared to the anti-Jovian side of the emission for each orbit.
The black triangles are calculated from the emission within the two
\ion{O}{1} disks only, while the red stars show the ratio calculated
from an extended disk, i.e. within a radius of 1.3 $R_E$.
The sub-Jovian/anti-Jovian emissions within the disk 
are within the error bars consistent with 
a ratio of one, i.e.
consistent with a symmetric emission. The extended disk emission
shows an average ratio of $\sim$0.9 \textcolor{black}{$\pm 0.2$.
Similar to UV observations of Io's atmospheric emission by 
\cite{roes99}, one might expect the anti-Jovian side of Europa
to be brighter due to the Hall effect \cite[]{saur99a,saur00a}.
However, this effect is not resolvable within the current 
error bars. In summary, no sub-Jovian and anti-Jovian asymmetry
on the leading side of Europa is visible in contrast to its
trailing side.}

\begin{figure}
\epsscale{1.6}
%\plotone{../../analysis9e/figures/fig_nimmo_psf_n08bj.eps}
\plotone{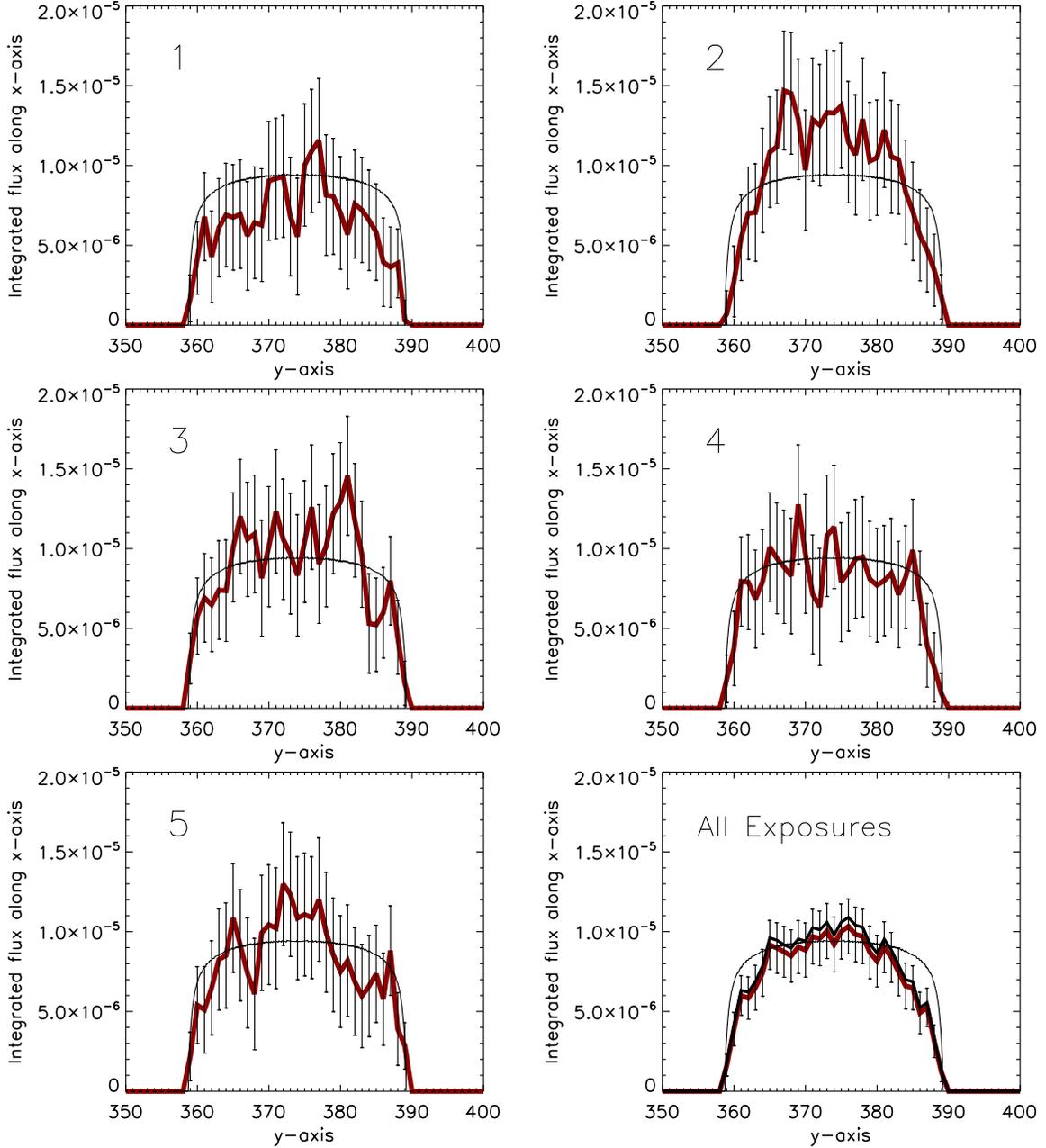}
\vspace*{0cm}
\caption{Integrated flux in the x-direction (north south) within
the disk of Europa plotted as a function
of the y-direction (sub-Jovian/anti-Jovian).  Red curve are observations and
thin black line is expected radiation from a radially symmetric atmosphere
including the point spread function. 
\textcolor{black}{
The black jagged line next to
the red one in the lower right panel is calculated
with the alternative model (see text). }
\label{f_nimmo}}
\end{figure}

In a next step we quantitatively search if there 
is substructure in the emission on the disk of the leading hemisphere.
Such a substructure might be generated by inhomogeneities of
Europa's surface properties such as discussed by \cite{cass07} or
plumes speculated by \cite{nimm07a} similar to those observed
on Enceladus \cite[e.g.,][]{wait06,spen06,porc06,doug06,hans06,burg07,saur08}.
Therefore we integrate the fluxes within the disk parallel to the
dispersion, i.e. along the x-axis. The resultant integrated
fluxes are shown as a function of the sub-Jovian/anti-Jovian direction 
as red curves plus error bars in thin black
for all individual orbits and the sum of all
exposures in Figure \ref{f_nimmo}. For comparison we
show as black curves the expected emissions from within the disk for a radially
symmetric atmosphere with a scale height of 145 km \cite[]{hall95,saur98}.
The effects of the point spread function on the radially symmetric model
atmosphere is included.
The model curve for a radially symmetric atmosphere is arbitrarily normalized
and displayed 
\textcolor{black}{
to demonstrate possible deviations from a radially
symmetric atmosphere in the observed
emission in the sub-Jovian/anti-Jovian direction.
The observations for the individual orbits show 
temporal variations in the overall flux. 
As described in Section \ref{ss_totalfluxes} 
overall temporal variations of the total intensity are consistent
with Europa's varying position with respect to the current sheet.
}

\textcolor{black}{
We use 
the sum of all exposures, which combines the \ion{O}{1} 1356 \A and
the \ion{O}{1} 1304 \A emission,
to investigate possible spatial
substructure within the emission.
There is more emission near the central meridian (90 degree
at eastern elongation) compared to the emission from near the
limb as the emission near the limb (near pixel 360 and 390)
is smaller compared to the emission profile expected from a radially
symmetric atmospheric emission. 
The observed emission could be consistent
with a density enhancement around the 90 degree meridian.
Only as an example to characterize the amount of gas, 
an atmospheric substructure 
with a radius of $\sim$250 km and a  density enhancement by a
factor of two to three  compared to the background atmosphere 
would create the observed deviations in Figure \ref{f_nimmo}
compared to a radially symmetric 
atmosphere. Such density variations could be consistent with
model calculations reported by \cite{cass07}.
In the right lower panel in Figure \ref{f_nimmo}, which
covers the sum of all exposures, we also show
the alternative model derived from an alternative throughput curve
as a jagged black line close to the jagged red line. It demonstrates
that the applied modifications in the throughput does not have
qualitative implications on our derived conclusions.
}
\textcolor{black}{
We also note that in Figure \ref{f_nimmo} lower right panel at least four
data points close to the limb
are within the error bars below the radially symmetric model. 
We could thus also combine two or all four to a joint point with even
smaller error bars. 
}

\begin{figure}
\epsscale{0.7}
%\plotone{../../analysis9b/figures/ojakangas.eps}
\plotone{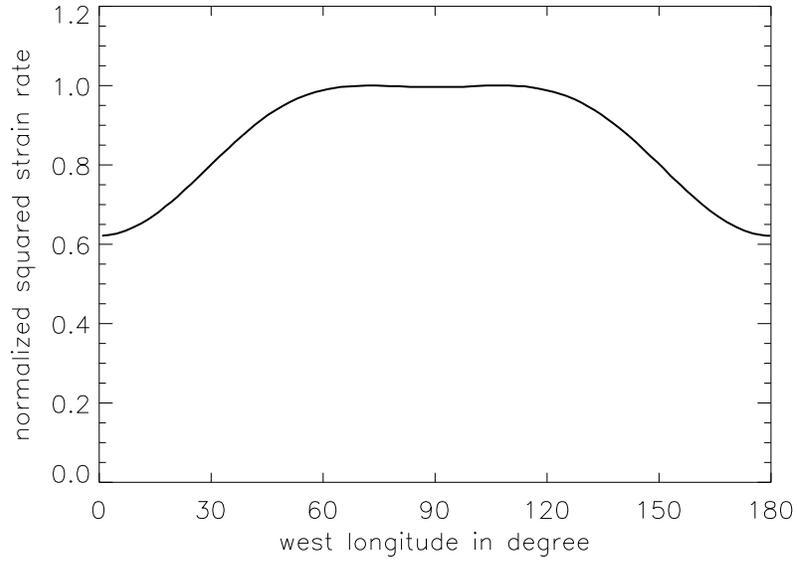}
\vspace*{0cm}
\caption{The time average of the squared tidal strain rate for a thin ice shell
\cite[see][Fig 1]{ojak89}, averaged over -90\grad to 90\grad latitude and plotted as a function of longitude. Values have been normalized so that the maximum value is 1. \label{f_ojakangas}}
\end{figure}

\cite{nimm07} suggested that the vapor plumes discovered at 
Enceladus were caused by tidally-driven shear heating, with the vapor 
production rate dependent on the local strain rate. If similar shear 
heating \cite[]{nimm02} and vapor production occurs at Europa, then 
the local rate of vapor production should depend on the local strain rate. 
For a thin ice shell, the time average of the squared tidal strain rate 
varies spatially, with maxima at the poles and equatorial minima at 
0\grad and 180\grad west 
longitude \cite[]{ojak89}. Figure \ref{f_ojakangas} 
plots the latitudinal average of this quantity as a function of 
longitude, demonstrating that strain rates (and hence predicted vapor 
production) are largest around 90\grad and 270\grad west longitude.
This analysis is probably over-simplified, in that it assumes the 
orientation and areal density of shear-heating features is uniform.
\textcolor{black}{
To take the spatial variation of shear-heating features into account, 
we also calculated the mean resolved shear stress (a proxy for shear 
heating rate) on mapped lineaments. The stress calculations are 
described in the Supplementary Information of \cite{nimm07}
%Nimmo et al. (2007b) 
and the lineament map was based on Figure 15b of \cite{katt09}.
%Doggett et al. (2009). 
Using this approach, we predict a peak in vapor production}
% at 230 W}
%For instance, by resolving shear stresses onto individual mapped 
%lineaments \citep[cf.][]{nimm07} we predict a peak in vapor production 
at 230\grad west, in striking agreement with the observations
\textcolor{black}{on the trailing side reported by \cite{mcgr04}
and shown in Figure \ref{f_western} of this paper}. 
However, the location of this peak may be an artifact due to the 
incomplete imaging coverage of polar regions \cite[Figure 1.]{dogg09}. 
If the observed peak at 230\grad west \cite[]{mcgr04,mcgr09}
is not a transient feature, and arises because 
of shear heating, then we predict that future imaging will reveal a 
paucity of polar ridges around 270\grad  west.  

The observed surpluses of \ion{O}{1} emission near 90\grad
and in between 200\grad and 250\grad west longitude need further
investigations. 
\textcolor{black}{
For example, if the surplus \ion{O}{1} emission were confined to 1304 \A 
and not 1356 \AA, it could indicate an optically thick atmosphere 
in O atoms, which would resonantly scatter solar 1304 \A photons 
with enhanced emission at the center of the disk.  \cite{hall95}
used the '2:1 ratio' on \ion{O}{1} 1356/\ion{O}{1} 1304 intensity 
to put an upper limit on the O atom column density of 
$< 2 \times 10^{18}$ m$^{-2}$.  But this ratio has been observed 
to be $<$ 2:1 (cf. \cite{hall98}) and 
thus solar 1304 \A resonance scattering by O atoms 
may be non-negligible.
The observed surpluses} also might be due to varying surface properties,
such as porosity or albedo.  The electrodynamic interaction
of Europa's atmosphere with Jupiter's magnetosphere will also
generate asymmetries in the interaction, however it generally
acts to enhance emission on the flanks in contrast to the 
enhanced  emission in the wake. These effects warrant further
investigations which are however beyond the scope of this observational
paper.

\section{Summary and Discussion}
\label{s_sum}
Europa's \ion{O}{1} 1304 \A and 1356 \A 
emissions on the leading side observed on 2008 June 29 with
ACS/SBC prism PR130L show total fluxes in the
range of 130 to 230 Rayleigh. These values are generally comparable with 
previous observations. They are slightly smaller 
than fluxes obtained with HST/STIS on the trailing side 
in the range of 210 to 370 Rayleigh \cite[]{mcgr04,mcgr09},
but they are in the range or somewhat larger than previous 
GHRS observation of 137 Rayleigh
on the leading side \cite[]{hall98}. 
The ACS/SBC and the HST/STIS observations
demonstrate a clear dependence of the emission
depending of Europa's position relative to
Jupiter's current sheet. We do not find an asymmetry
of the atmospheric emission with respect to the 
sub-Jovian/anti-Jovian side, but find a surplus of emission near 
90\grad west longitude compared to a radially symmetric atmosphere.
The previous STIS observations are consistent with a
surplus of emission on the trailing side in the range of
200\grad to 250\grad west longitude \cite[]{mcgr04,mcgr09}. 
A surplus of emission could be due to a surplus of atmospheric gas, which 
could be consistent with predictions of possible 
plume locations on Europa by \cite{nimm07a} near the same
longitudes.
Other reasons which could contribute to the observed
longitudinal enhancement in the UV emission might be 
intrinsic or solar illumination driven 
spatial variations of Europa's ice properties
controlling its albedo, porosity and sputtering properties
\cite[e.g.,][]{cass07}. % possibly due inhomogeneous solar illumination.
For example, the albedo is known to vary with longitude on Ganymede for
wavelengths larger than 2200 \A \cite[]{nels87,noll95,hend98},
however it appears to be constant within the error bars in the 
FUV at 1335 \A \cite[]{hall98}.
The emission patterns of Europa's atmosphere are also controlled
by the plasma interaction with Jupiter's magnetosphere. Europa's
\ion{O}{1} 1304 \A and \ion{O}{1} 1356 \A emissions are generated by electron impact
dissociation and are thus a strong function of electron density and
temperature. Europa's leading side is the downstream side of the
plasma convection past Europa. Since Europa absorbs the plasma flow
including the electrons on its upstream side 
a rarefied wake of electrons is generated on the downstream, i.e.
on the leading side. Additionally, the electron temperatures
are decreased compared to the upstream side due to inelastic
collisions with Europa's atmosphere. These effect however
rather predicts smaller emission near the 90\grad west-longitude
instead of the observed surplus, but it 
naturally explains the observed 
smaller emission from the leading side compared
to the trailing disk of Europa.

Europa's UV emissions are intriguingly puzzling. The limitations
caused by the HST/ACS red leak problem and its associated data analysis
problems require further 
observations  with HST/STIS
at various orbital longitudes 
and in addition measurements by a possible future 
Europa orbiter  to better understand Europa's 
atmospheric emission 
and its hemispheric differences. Also further numerical simulations
are needed to investigate how plumes combined with the time variable
plasma interaction possibly control the temporal evolution of Europa's UV
emissions.

\appendix

\section{Appendix: Calibration of prism properties }
\begin{figure}
\epsscale{1.}
%\plotone{../../analysis9e/figures/16Cygb_n08bj.eps}
\plotone{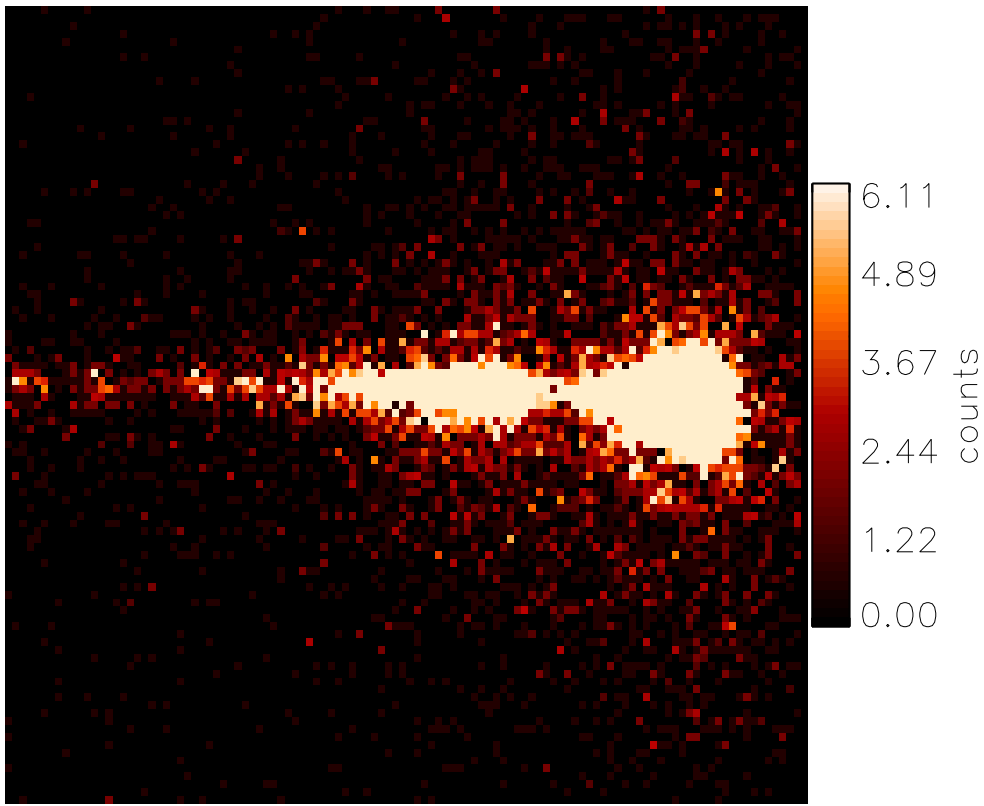}
\vspace*{0cm}
\caption{Observations of 16CygB. \label{f_obs_star}}
\end{figure}

For the ACS/SBC prism PRL130, both the throughput for wavelengths larger
than 2000 \A (the red leak) and the point spread function are not
available through the calibration pipeline, so we constrain them
here. We estimate both properties with 
observations of 16 CygB, a solar analog star.
16 CygB was observed with ACS/SBC PR130L in Cycle 16 (ID 11325, 
PI Kuntschner) to improve the existing sensitivity calibrations.
In Figure \ref{f_obs_star}
we show observations of 16 CygB dispersed with PR130L. 
16 CygB is not a monochromatic point source. In case
there were no scattering, the dispersed point source would thus lead
to a 'line of emission' on the detector along the dispersion without 
spread perpendicular to it (i.e. in our case without spread in the vertical
direction in Figure \ref{f_obs_star}). The dispersion
and thus the 'line of emission' is
oriented predominantly in the horizontal direction
in Figure \ref{f_obs_star} (similar to our Europa observations). 
The scattering additionally spreads
the emission also to perpendicular to the 'line of emission'.
The horizontal direction is also called x-direction and
vertical direction the y-direction.

\subsection{ Long wavelength throughput}
\label{ss_throughput}
\begin{figure}
\epsscale{1.1}
%\plotone{../../analysis9e/figures/throughput_from_cygb_feld3h.eps}
\plotone{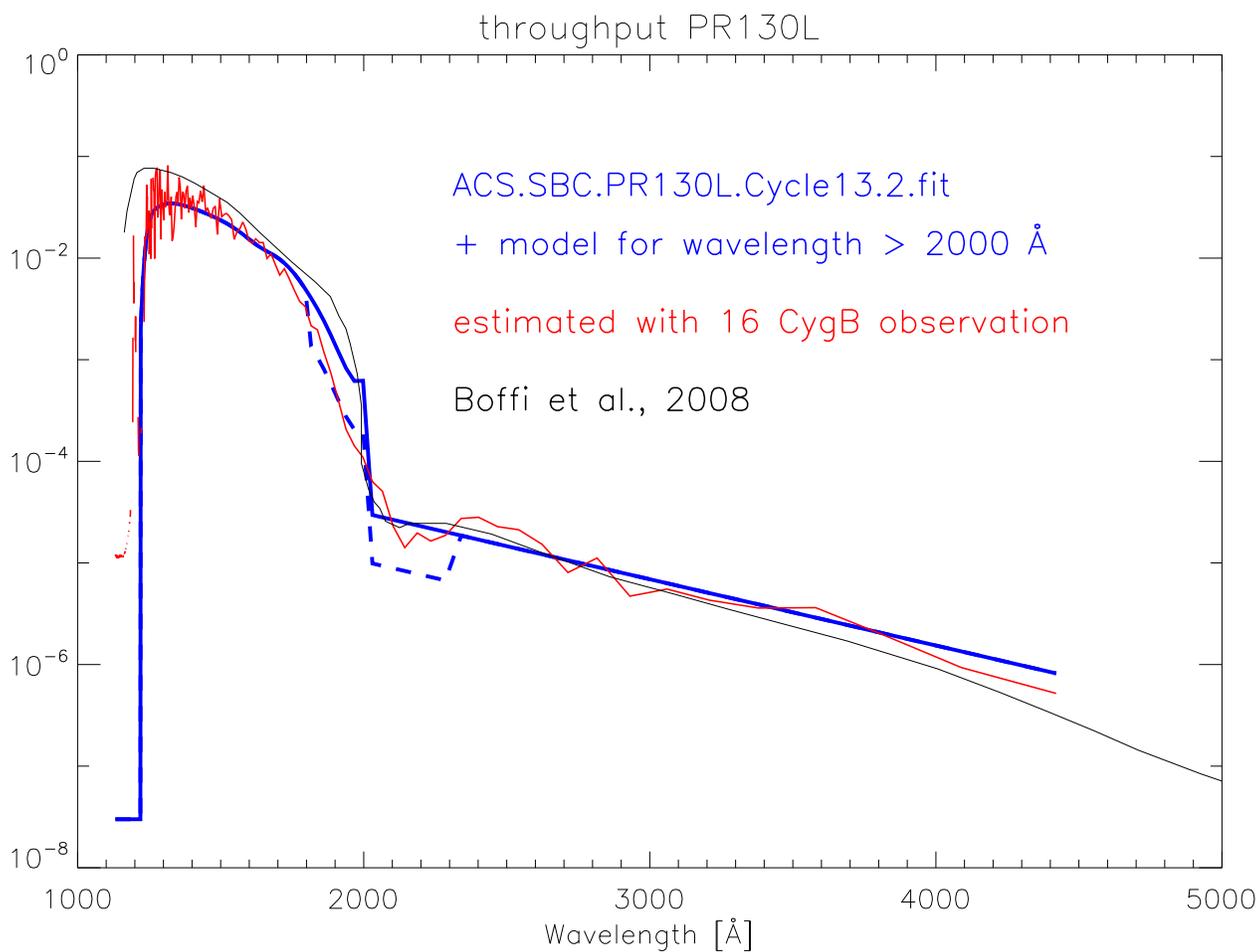}
\vspace*{0cm}
\caption{Throughput of the prism PR130L. Blue curve is the calibrated
throughput for wavelengths shorter than
2000 \A from \cite{lars06} 
and modeled with an exponential law for wavelengths larger
than 2000 \A to fit 16 CygB observations. 
\textcolor{black}{
An alternative model for the throughput is used in the paper with
a reduced throughput by factor 1/3 within 1800 \A and 2300 \A 
(blue dashed line). 
}
Red curve is calculated throughput
from 16 CygB observations with solar analog spectrum.
We additionally show the total ACS/SBC throughput derived for Synphot by
\cite{boff08} in their Figure 3 as thin black line. 
\label{f_throughput}}
\end{figure}

The throughput of the prism PR130L 
for wavelengths larger than 2000 \A 
can be estimated with the 16 CygB observations. 
We employ a solar analog spectrum
\cite[]{wood96,neck94,arve69,coli96}
which we disperse as a point source using the dispersion
described with formula (\ref{e_model}).
To retrieve an estimate for the throughput, 
we integrate the 16 CygB observations along the y-direction on the 
detector, i.e. collapse the 2D emission onto the 'line of emission'
and divide it by the modeled emission of the solar analog star.
The throughput is displayed as a function of wavelength in
Figure \ref{f_throughput} as the red curve. It agrees fairly
well with the calibrated throughput of the prism 
PR130L for wavelengths
shorter than 2000 \A shown in blue. For wavelengths larger
than 2000 \AA, our derived throughput serves as basis
to construct a model throughput. For $\lambda>$ 2000 \A we
therefore simply use an exponential law which starts at 2000 \A with
a value of a factor of 20 smaller than the calibrated throughput
value at $\lambda$ just lower than 2000 \AA. It
reads
\begin{eqnarray}
t(\lambda)=t_0 \exp\left[-\frac{\lambda-\lambda_0}{H_{red}}\right]
\end{eqnarray}
with $t_0$ = 3$\times$10$^{-5}$, $\lambda_0$ = 2000 \A and $H_{red}$ = 2000/3 
\AA.
 Our model throughput
for $\lambda>$ 2000 \A (blue curve) matches the throughput
estimated from the 16 CygB observations (red curve) fairly well.
In Figure \ref{f_throughput} we also show for comparison
the total ACS/SBC throughput
rederived by \cite{boff08} for Synphot.
In summary, for $\lambda <$ 2000 \A we use the throughput derived
by \cite{bohl00} and for  $\lambda >$ 2000 \A we use as 
model throughput an exponential law (blue curve in Figure \ref{f_throughput}).

\textcolor{black}{
We also apply a modified, i.e. alternative, throughput in our analysis to 
study systematic uncertainties in the calculation of the surface
reflected light. Therefore we assume a throughput dip in the
wavelengths range 1800 \A to 2300 \A by factor of 1/3 applied to the model
throughput derived in the previous paragraph. The alternative throughput
curve is shown as dashed blue line in Figure \ref{f_throughput}.
}

\subsection{ Point Spread Function}
\label{ss_psf}

\begin{figure}
\epsscale{1.1}
%\plotone{../../analysis9e/figures/psf_n04cn.eps}
\plotone{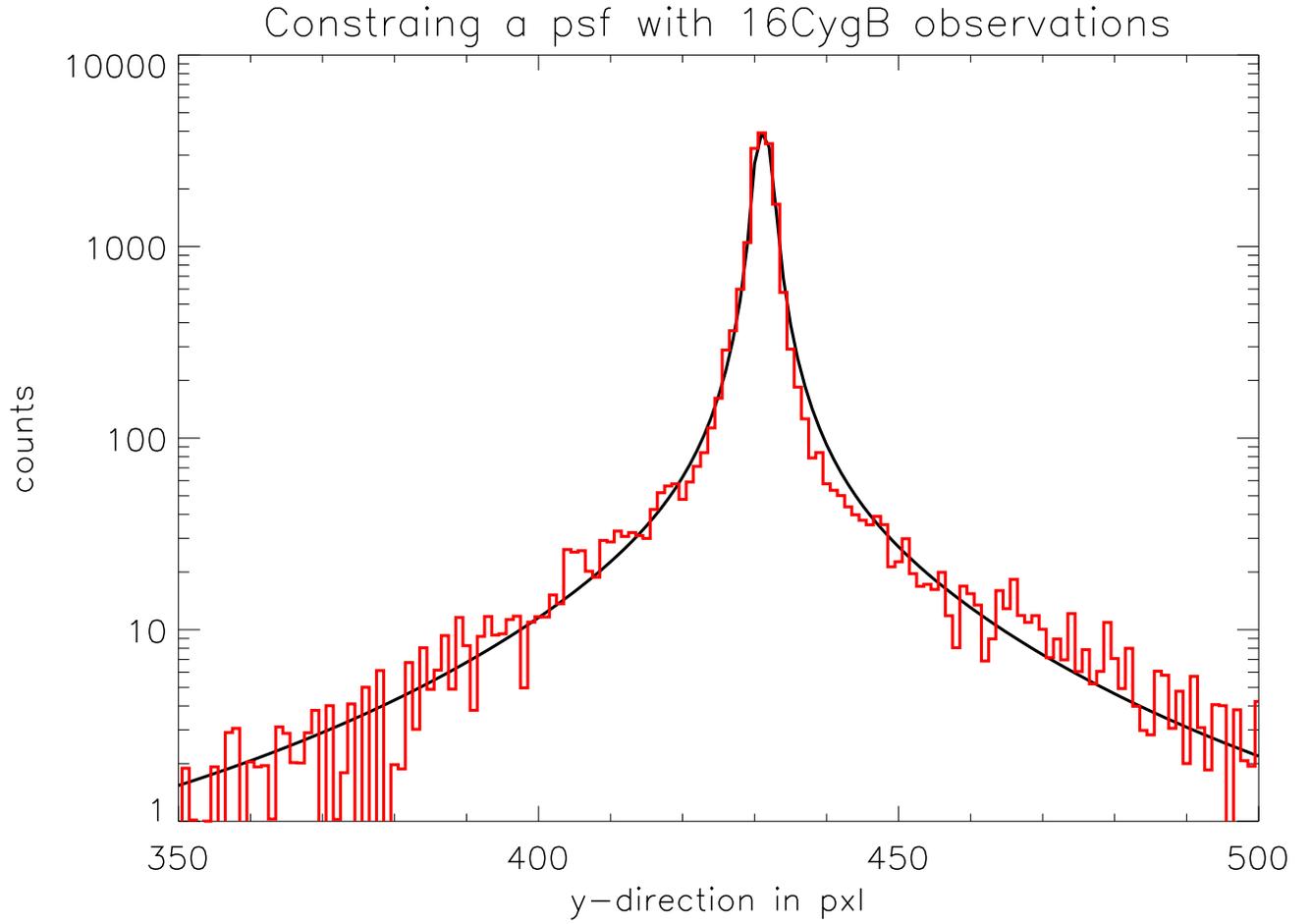}
\vspace*{0cm}
\caption{Estimation of the point spread function (psf) in y-direction.
Observations of 16CygB in red and model flux to constrain the
psf in black. \label{f_psf}}
\end{figure}

We use the 16 CygB observations (ID 11325, 
PI Kuntschner) also to construct a point spread function (psf)
to be used for the Europa observation. 
Observations of 16 CygB
with prism PR130L  would lead
to a 'line of emission' on the detector along the dispersion in
case there is no scattering. Occurring scattering 
spreads this line. The scattering 
is quantified in Figure \ref{f_psf}, where we show the observations 
of 16 Cyg B integrated
parallel to the dispersion in the x-direction
(columns 268 to 304) as a function
of rows (y-direction) as red curve.
The observations maximize near row y=432. Near y=432 the flux
might be described as a Gaussian, but further away it 
does not show obvious exponential decay and rather resembles
a power law. 

To model the observed scatter shown in Figure \ref{f_psf}, we
use the solar analog spectrum, disperse it with the prism PR130L
properties as detailed in (\ref{e_model}), and then convolve it
with a point spread function to be determined.
The scattering 
as shown in Figure \ref{f_psf} can be modeled
by  a point spread function of the form
\begin{eqnarray} 
\mbox{psf}(x,y)=c_0 \left[1 + \left(\frac{r}{w}\right)^{\gamma}\right]^{-\kappa/\gamma}
\label{e_psf}
\end{eqnarray} 
with $r=\sqrt{x^2+y ^2}$. The mathematical expression in (\ref{e_psf})
has three free parameters. 
The first parameter $w$ describes the width of 
the structure. Its value is related to the full width at half maximum
(fwhm) by fwhm $\approx w \times 2^{1+1/\kappa}$ when $\gamma \gg \kappa$. 
The second parameter
$\gamma$ describes the shape of the
function around its maximum and the third parameter
$\kappa$ models a power law decay further from the maximum. The coefficient
$c_0$ is introduced such  
that the two-dimensional integral over the psf renders unity, i.e.
$\int\int \mbox{psf}(x,y)\; dx\; dy=1$.
This psf resembles a modified version of the $\kappa$-distribution,
a well known phase space density function in plasma physics.
The $\kappa$-distribution approaches a Gaussian
for large $\kappa$ and describes a power law for small $\kappa$.
The psf in (\ref{e_psf}) turns into an interesting 
sub-class of point spread functions 
for $\gamma$=2, and $w =\sigma \sqrt{\kappa}$. In this case (\ref{e_psf})
leads to 
%$\mbox{psf}(x,y)=c_0\left[1+(r/(\sigma \sqrt{\kappa})^2)\right]^{-\kappa/2}$.
\begin{eqnarray}
\mbox{psf}(x,y)=c_0\left[1+(\frac{r}{\sigma \sqrt{\kappa}})^2)\right]^{-\kappa/2} \;.
\end{eqnarray}
This psf resembles for finite $\kappa$ 
a power law with exponent $-\kappa$ for sufficient
distances from the source. It turns for large $\kappa$, i.e. in the limit 
$\kappa \rightarrow \infty$ to a Gaussian with variance $\sigma^2$. 

We convolve the psf$(x,y)$ in (\ref{e_psf}) with the line source calculated with
the solar analog spectrum discussed above: 
The coefficients in the psf are constrained by
the integrated 16 CygB measurements shown in Figure \ref{f_psf}. 
We determine the coefficients of
the psf quantitatively by fitting it (black curve) 
to the integrated fluxes along
the dispersion (red curve). 
We find a good fit for $\kappa$ = 2.65, $w$ = 1.17 and $\gamma=20$.
For these parameters the derived psf reproduces 
the 16 CygB observations well, both near the maximum and
along the wings (see Figure \ref{f_psf}).

\acknowledgments

This work is based on observations with the NASA/ESA {\it Hubble
Space Telescope} obtained at the Space Telescope Science Institute,
which is operated by the Association of Universities for Research in
Astronomy (AURA), Inc., under NASA contract NAS 5-26555. 
We thank A. Roman for scheduling the observations. J.S. appreciates the
hospitality of the Johns Hopkins University during his sabbatical stay in
spring/summer 2011. 
L.R. and J.S. acknowledge support by the Verbundforschung Astronomie
und Astrophysik durch das Bundesministerium fuer Wirtschaft und Technologie.
 P.D.F., D.F.S., and K.D.R.
were supported by NASA grant HST-GO-11186.01-A. 
J.C.G. and D.G. are supported by the Belgian Fund for Scientific Research (FNRS)and by a PRODEX contract with the European Space Agency, managed by the Belgian Federal Space Policy Office.

\bibliographystyle{apj}
\bibliography{../../../../literatur/lit}

\clearpage

%% the file, uncomment the \plotone, \plottwo, and
%% If you need a layout that cannot be achieved with \plotone or
%% \plottwo, you can invoke the graphicx package directly with the
%% \includegraphics command or use \plotfiddle. For more information,
%% please see the tutorial on "Using Electronic Art with AASTeX" in the
%% documentation section at the AASTeX Web site,
%% http://www.journals.uchicago.edu/AAS/AASTeX.
%%
%% The examples below also include sample markup for submission of
%% supplemental electronic materials. As always, be sure to check
%% the instructions to authors for the journal you are submitting to
%% for specific submissions guidelines as they vary from

%% This example uses \plotone to include an EPS file scaled to
%% 80% of its natural size with \epsscale. Its caption
%% has been written to indicate that additional figure parts will be
%% available in the electronic journal.

\clearpage

\clearpage

\clearpage

\clearpage
\clearpage

\clearpage
\clearpage

\clearpage
\clearpage

\clearpage

\clearpage

\clearpage

\clearpage

\clearpage

\clearpage

\clearpage

\clearpage

\clearpage

\clearpage
\clearpage

\begin{table}
\begin{center}
\caption{Properties of HST program 11186 executed on 2008-06-29\label{t_1}}
\begin{tabular}{ccccccccc}
\tableline\tableline
Orbit & Start time & Exposures & Duration\tablenotemark{a}  & $\alpha$\tablenotemark{b}  & Total Flux\tablenotemark{c} &Total Flux\tablenotemark{c}  & \\
 &in UT & & in s& in degree  & in 10$^{-5}$ photons s$^{-1}$ cm$^{-2}$ & in Rayleigh &
\multicolumn{1}{c}- \\
%\multicolumn{1}{c}{$\Theta$\tablenotemark{b}} \\
\tableline
1 &04:13:14  &4 &641  &-9.1 &21 $\pm$ 2.1 &132 $\pm$ 14\\
2 &05:48:16  &4 &641 &-1.9  &34 $\pm$ 2.2 &216 $\pm$ 14\\
3 &07:24:12 &4  &641  &6.1  &36 $\pm$ 2.2 &226 $\pm$ 14\\
4 &09:00:07 &4 &641  &9.6   &29 $\pm$ 2.2 &187 $\pm$ 14\\
5 &10:36:02 &4 &652 &8.0    &32 $\pm$ 2.2 &202 $\pm$ 14\\
\tableline
\end{tabular}
\tablenotetext{a}{per exposure in each orbit}
\tablenotetext{b}{Nominal angle of Europa's position with 
respect to the magnetic equator.}
\tablenotetext{c}{\ion{O}{1} 1304  \AA \hspace{+0.1mm} and \ion{O}{1} 1356  \AA \hspace{+0.1mm}
combined. The error bars include statistical errors only.}
\end{center}
%\label{t_1}
\end{table}

%% If the table is more than one page long, the width of the table can vary
%% from page to page when the default \tablewidth is used, as below.  The
%% individual table widths for each page will be written to the log file; a
%% maximum tablewidth for the table can be computed from these values.
%% The \tablewidth argument can then be reset and the file reprocessed, so
%% that the table is of uniform width throughout. Try getting the widths
%% from the log file and changing the \tablewidth parameter to see how
%% adjusting this value affects table formatting.

%% The \dataset{} macro has also been applied to a few of the objects to
%% show how many observations can be tagged in a table.

\clearpage

\end{document}